\documentstyle[psfig]{mn}

\begin{document}

\journal{To Appear in MNRAS}

\title[X-ray isophote shapes and the mass of NGC 3923]
{X-ray isophote shapes and the mass of NGC 3923}
\author[D.~A. Buote and C.~R. Canizares]{David
A. Buote$^1$ and Claude R. Canizares$^2$ \\ 
$^1$Institute of Astronomy, Madingley Road, Cambridge CB3 0HA \\
$^2$Department of Physics and Center for Space Research 37-241,
Massachusetts Institute of Technology,  
77 Massachusetts Avenue, \\ Cambridge, MA 02139, U.S.A.}

\maketitle

\begin{abstract}

We present analysis of the shape and radial mass distribution of the
E4 galaxy NGC 3923 using archival X-ray data from the {\sl ROSAT} PSPC
and HRI. The X-ray isophotes are significantly elongated with
ellipticity $\epsilon_x=0.15 (0.09-0.21)$ (90\% confidence) for
semi-major axis $a\sim 10h^{-1}_{70}$ kpc and have position angles
aligned with the optical isophotes within the estimated
uncertainties. Applying the Geometric Test for dark matter, which is
independent of the gas temperature profile, we find that the
ellipticities of the PSPC isophotes exceed those predicted if
$M\propto L$ at a marginal significance level of $85\% (80\%)$ for oblate
(prolate) symmetry. Detailed hydrostatic models of an isothermal gas
yield ellipticities for the gravitating matter,
$\epsilon_{mass}=0.35-0.66$ (90\% confidence), which exceed the
intensity weighted ellipticity of the $R$-band optical light, $\langle
\epsilon_R\rangle = 0.30$ ($\epsilon_R^{max}=0.39$).

We conclude that mass density profiles with $\rho\sim r^{-2}$ are
favored over steeper profiles if the gas is essentially isothermal
(which is suggested by the PSPC spectrum) and the surface brightness
in the central regions $(r\la 15\arcsec)$ is not modified
substantially by a multi-phase cooling flow, magnetic fields, or
discrete sources.  We argue that these effects are unlikely to be
important for NGC 3923.  (The derived $\epsilon_{mass}$ range is very
insensitive to these issues.)  Our spatial analysis also indicates
that the allowed contribution to the {\it ROSAT} emission from a
population of discrete sources with $\Sigma_x\propto\Sigma_R$ is
significantly less than that indicated by the hard spectral component
measured by {\sl ASCA}.

\end{abstract}

 
\section{Introduction}
\label{intro}

The structure of the dark matter halos of galaxies provides important
clues to their formation and dynamical evolution (e.g.  Sackett 1996;
de Zeeuw 1996, 1997). For example, in the Cold Dark Matter (CDM)
scenario (e.g.  Ostriker 1993) there is evidence that the density
profiles of halos have a universal form essentially independent of the
halo mass or $\Omega_0$ (Navarro, Frenk, \& White 1997; though see
Moore et al. 1997). The intrinsic shapes of CDM halos are
oblate-triaxial with ellipticities similar to the optical isophotes of
elliptical galaxies (e.g.  Dubinski 1994). The global shape of a halo
also has implications for the mass of a central black hole (e.g.
Merritt \& Quinlan 1997).

At present accurate constraints on the intrinsic shapes and density
profiles of early-type galaxies are not widely available (e.g. 
Sackett 1996; Olling \& Merrifield 1997)\footnote{The distribution of
dark matter in spiral galaxies is also far from being a solved problem
-- see, e.g.  Broeils \shortcite{broeils}.}. Stellar dynamical
analyses that have incorporated the information contained in high
order moments of stellar velocity profiles have made important
progress in limiting the uncertainty in the radial distribution of
gravitating mass arising from velocity dispersion anisotropy (Rix et
al. 1997; Gerhard et al. 1997). However, as indicated by the paucity
of such stellar dynamical measurements, the required observations to
obtain precise constraints at radii larger than $\sim R_e$ are
extensive, and the modeling techniques to recover the phase-space
distribution function are complex. It is also unclear whether this
method can provide interesting constraints on the intrinsic shapes
since only weak limits on the range of possible shapes have been
obtained from analysis of velocity profiles out to $\sim 2$ $R_e$
(e.g.  Statler 1994).

Interesting measurements of the ellipticity of the gravitating mass
have been obtained for two Polar Ring galaxies (Sackett et al. 1994;
Sackett \& Pogge 1995) and from statistical averaging of known
gravitational lenses (e.g.  Keeton, Kochanek, \& Falco 1997), but
owing to the rarity of these objects it is possible that the
structures of their halos are not representative of most early-type
galaxies. Moreover, gravitational lenses, which are biased towards the
most massive galaxies, only give relatively crude constraints on the
ellipticity and radial mass distribution for any individual system and
only on scales similar to the Einstein radius (e.g.  Kochanek 1991).

The X-ray emission from hot gas in isolated early-type galaxies
(Forman, Jones, \& Tucker 1985; Trinchieri, Fabbiano, \& Canizares
1986; for a review see Sarazin 1997) probably affords the best means
for measuring the shapes and radial mass distributions in these
systems (for a review see Buote \& Canizares 1997b; also see Schechter
1987 and the original application to the analogous problem of the
shapes of galaxy clusters by Binney \& Strimple 1978). The isotropic
pressure tensor of the hot gas in early-type galaxies greatly
simplifies measurement of the mass distribution over stellar dynamical
methods. Moreover, since the shape of the volume X-ray emission traces
the shape of the gravitational potential independent of the (typically
uncertain) gas temperature profile (Buote \& Canizares 1994, 1996a),
the shape of the mass distribution can be accurately measured in a way
that is quite robust to the possible complicating effects of
multi-phase cooling flows and magnetic fields (see Buote \& Canizares
1997b).

Presently, X-ray measurements of the mass distributions in early-type
galaxies are inhibited by limitations in the available data. The {\sl
ROSAT} \cite{trump} Position Sensitive Proportional Counter (PSPC)
\cite{pf} has inadequate spatial resolution (PSF $\sim 30\arcsec$
FWHM) to map the detailed mass distributions for all but the largest
nearby galaxies, and the limited spectral resolution and band width
complicates interpretation of the measured temperature profiles (Buote
\& Canizares 1994; Trinchieri et al. 1994; Buote \& Fabian
1997). Although equipped with superior spatial resolution (PSF $\sim
4\arcsec$ FWHM), the {\sl ROSAT} High Resolution Imager (HRI)
\cite{david} has too small an effective area and too large an internal
background to provide images of sufficient quality for many galaxies
for radii $r\ga R_e$. Among the few galaxies with detailed
measurements of their radial mass profiles are NGC 507 (Kim \&
Fabbiano 1995), NGC 1399 (Rangarajan et al. 1995; Jones et al. 1997),
NGC 4472 (Irwin \& Sarazin 1996), NGC 4636 (Trinchieri et al. 1994),
NGC 4649 \cite{bm}, and NGC 5044 (David et al. 1994).

The shape of the gravitating mass has been measured via X-ray analysis
for the E4 galaxy NGC 720 and the E7/S0 galaxy NGC 1332 and found to
be at least as elongated as the optical isophotes (Buote \& Canizares
1994, 1996a, 1997a).  For NGC 720, which has more precise constraints,
the ellipticity of the gravitating matter is
$\epsilon_{mass}=0.44-0.68$ (90\% confidence) compared to the
intensity weighted ellipticity of the optical light,
$\langle\epsilon\rangle=0.31$ (Buote \& Canizares 1997a).  In
addition, the X-ray isophotes of NGC 720 twist from being aligned with
the optical isophotes within $R_e$ to a position $\sim 30\degr$ offset
at larger radii. This twist, when combined with the ellipticities of
the X-ray isophotes, cannot be explained by the projection of a
reasonable triaxial matter distribution and thus may implicate a dark
matter halo misaligned from the stars (Buote \& Canizares 1996b;
Romanowsky \& Kochanek 1997).

NGC 720 and NGC 1332 were selected for analysis since they are
isolated, significantly elongated in the optical, sufficiently bright,
and sufficiently dominated by emission from hot gas in the {\sl ROSAT}
band.  In this paper we present X-ray analysis of the classic
``shell'' galaxy, NGC 3923, which is the last galaxy of which we are
aware that satisfies these selection criteria and has deep {\sl ROSAT}
observations.  This isolated E4 galaxy has both archival {\sl ROSAT}
PSPC and HRI data and its {\sl ASCA} spectrum has been analyzed
previously \cite{bf}. This will serve as our final case study until
the impending launch of {\sl AXAF} revolutionizes this field.

The organization of this paper is as follows. In \S \ref{obs} we
describe the {\sl ROSAT} observations and the data reduction. We
discuss removal of point sources in \S \ref{pt}. Measurements of the
ellipticities of the X-ray isophotes and the radial profiles are
described in \S \ref{e0} and \S \ref{radpro} respectively. Analysis of
the PSPC spectrum is presented in \S \ref{spectra}. We give results
for the Geometric Test for dark matter in \S \ref{gt} and constraints
on the shape and radial mass distribution from detailed hydrostatic
models in \S \ref{models}. Finally, in \S \ref{conc} we give our
conclusions.

\section{Observations and data reduction}
\label{obs}

\subsection{PSPC}
\label{pspc}

NGC 3923 was observed with the PSPC for 14.5 ks from 17-19 December
1991 and for 24 ks from 23-26 June 1993. Both observations were
positioned at the field center where the point spread function is
smallest. Since spatial resolution is of principal importance for our
analysis of the shape of the X-ray surface brightness, we analyze only
the PSPC data in the hard band (PI channels 42-201, $E\approx 0.4-2.0$
keV) to further optimize the size of the PSF; for details regarding the
PSPC PSF see Hasinger et al. (1993, updated 1994) and see Aschenbach
\shortcite{asch} for a description of the {\sl ROSAT} X-ray
telescope.

We reduced each observation separately with the standard XSELECT,
FTOOLS, and IRAF-PROS software. Firstly, the events files were cleaned
of after-pulse signals by removing any events following within 0.35 ms
of a precursor. We then removed large fluctuations in the light curves
indicative of scattered light from the Bright Earth, Sun, or SAA; this
resulted in filtered exposures of 13.5 ks for the 1991 observation and
21.8 ks for the 1993 observation.

\begin{figure*}
\parbox{0.49\textwidth}{
\centerline{\psfig{figure=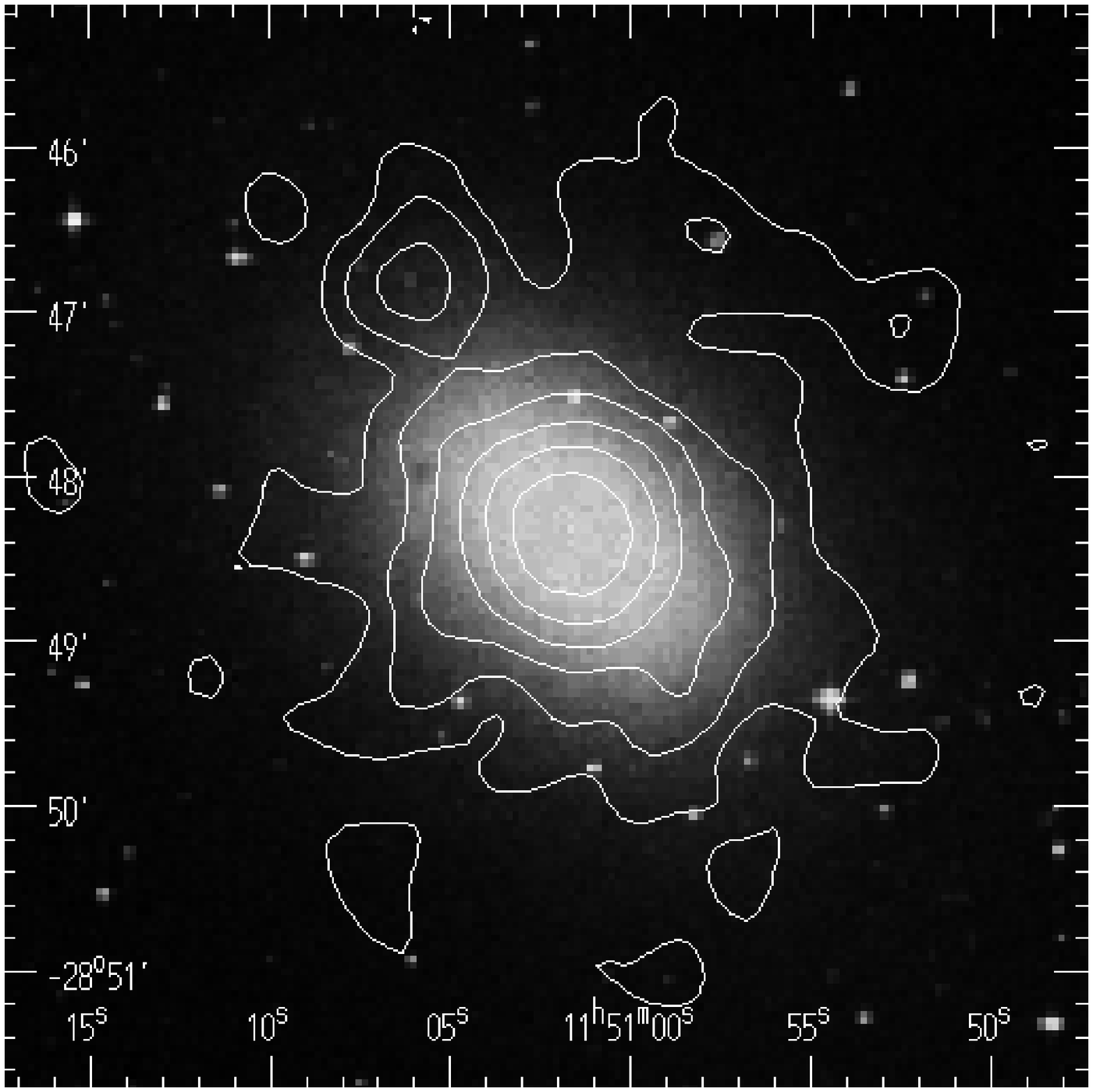,angle=0,height=0.3\textheight}}
}
\parbox{0.49\textwidth}{
\centerline{\psfig{figure=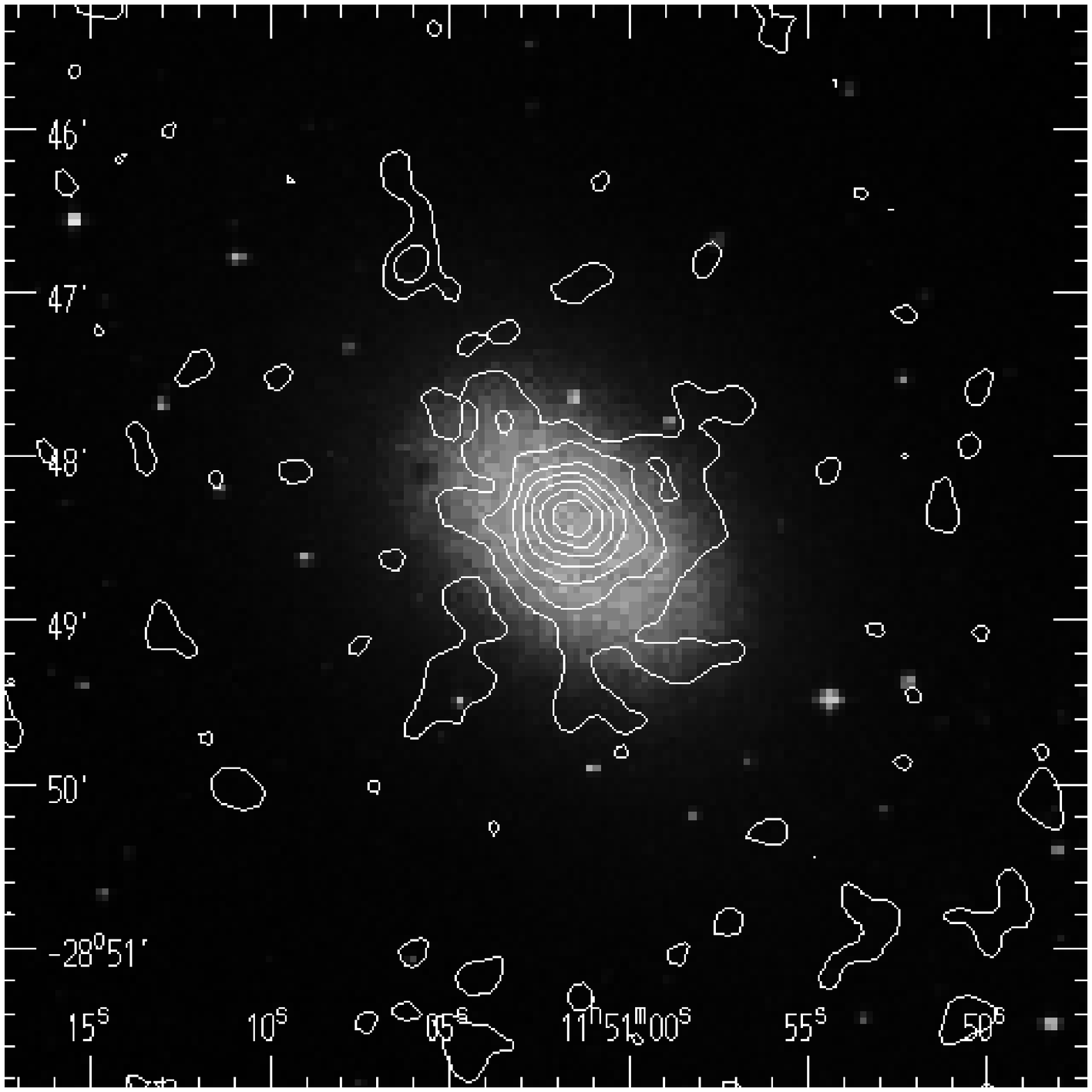,angle=0,height=0.3\textheight}}
}
\caption{\label{fig.n3923} Reduced images of the PSPC (left) and HRI
(right) observations overlaid on the digitized POSS images. Each X-ray
image has been smoothed with a Gaussian $(\sigma = 5\arcsec)$ for
visual clarity, although the images used for analysis have not been
smoothed as such.}
\end{figure*}

To optimize signal-to-noise (S/N) and bin-size requirements for
computing ellipticities (see \S \ref{e0}), we binned the cleaned
events files into images with $5\arcsec$ pixels. Exposure maps for the
appropriate energy band (PI=42-201) were then generated for each image
and were then used to create flattened images; note that the intrinsic
resolution of the exposure maps is $15\arcsec$. Finally, we aligned
the images using bright point sources in the field and then combined
the images. The final image is displayed in Figure \ref{fig.n3923}

\subsection{HRI}
\label{hri}

NGC 3923 was observed with the HRI for 45 ks from 10-17 January 1995
and for 26 ks from 22-26 June 1995. Both observations were positioned
at the field center where the PSF is smallest (David et al. 1997).
For detailed explanation of our reduction of HRI data we refer the
reader to the related study of NGC 720 by Buote \& Canizares
\shortcite{bc96b}.

We restricted the data to those in pulse-height analyzer (PHA) bins
2-8 since they maximized the S/N of the data of each observation. For
each observation we binned the events into images with $1\arcsec$
pixels to optimize determination of ellipticities within a $30\arcsec$
radius of the galaxy center (\S \ref{e0}). Exposure maps were
generated with the standard software for each image and then used to
flatten the images.

An accurate aspect solution is critical to analysis of isophote shapes
with the HRI on small scales $(r\la 15\arcsec)$, with the amplitude of
the asymmetrical distortion due to incomplete aspect correction
typically being most important for $(r\sim 5\arcsec - 10\arcsec)$
\cite{david}. The S/N of each of the NGC 3923 observations are too
small to usefully perform the aspect correction algorithm of Morse
\shortcite{morse}. The low S/N also makes it difficult to make strong
statements about the ellipticity of isolated point sources indicative
of aspect error. As a result, following Buote \& Canizares
\shortcite{bc96b} we search for possible aspect errors by examining
the positions of point sources for each of the OBIs; i.e. the time
intervals during which the spacecraft continuously pointed on the
target.

Three OBIs contain most of the ``on time'' of the January observation:
19 ks for OBI-6, 20 ks for OBI-11, and 9 ks for OBI-14; 28 ks is
distributed fairly evenly among 12 other OBIs. Using OBI-11 as a
reference, the bright point sources in the NGC 3923 field are
displaced by $2\arcsec\pm 1\arcsec$ for OBI-6, and $3\arcsec\pm
1.5\arcsec$ for OBI-14. These shifts, though statistically
significant, are very consistent with the expected aspect
uncertainties for a typical observation. To register all of the OBIs
to one coordinate frame, we binned together those OBIs that were
observed close together in time; i.e. OBIs 1-5, 6-9, 10-11, 12-13, and
14-15. Once registered to the same coordinate frame, the images were
all added. (Note that the image for each of these OBI groups was
binned as above.)

The June observation proved to be problematic because none of the 19
OBIs had a long enough exposure to provide very accurate source
positions. Hence, we were unable to perform a reliable test of the
aspect solution. In order to make some estimate of the error, we
grouped those OBIs closest together in time: OBIs 1-3, 4-12, 13-16,
and 17-19.  For three of these groups, less than $1\arcsec$ shifts
were required, but the OBI 4-12 group required an $8\arcsec\pm
1\arcsec$ shift. This large shift is questionable since all but one
source was too faint to even obtain a centroid measurement using the
alignment software in IRAF.

As it is unclear from considerations of the OBIs how accurate is the
aspect error for the June observation, we compared the ellipticity and
orientations (computed as described in \S \ref{e0}) of the surface
brightness with those of the January observation. Unfortunately, we
find significant disagreement between the two observations; note the
disagreement occurs whether or not we include the OBIs 4-12 for the
June observation. In particular, the position angles determined from
the January observation are fully consistent with the PSPC data and
the optical position angles for $r\la 30\arcsec$, which agrees with
NGC 720 and NGC 1332 (Buote \& Canizares 1996a,b); the ellipticities
are also consistent with the PSPC data.  The June observations,
however, have position angles that differ by over $60\degr$ and the
ellipticities, particularly for $r\ga 20\arcsec - 30\arcsec$ are less
than 0.1 as opposed to $\sim 0.25$ for the January
observation. Furthermore, the radial profile (see \S \ref{radpro}) of
the June observation is significantly flatter for $r\la 10\arcsec$.

Thus, the isophote shapes and orientations are inconsistent for the
two observations. The differences point to a serious aspect error in
the June observations. That is, although we would expect aspect error
to induce ellipticity in an intrinsically circular source, if the
distortion occurs nearly along the minor axis of a moderately
elliptical source, it will {\it reduce} the ellipticity of the source,
as well as alter the position angle. The flatter inner radial profile
of the June data is also consistent with uncorrected wobble. As a
result, we do not include the June data in our analysis. (This is not
critical since the S/N to be gained over the 45 ks is marginal.)

We display the reduced January observation in Figure \ref{fig.n3923}. 

\section{Point sources}
\label{pt}

It is readily apparent from inspection of Figure \ref{fig.n3923},
particularly for the higher S/N PSPC data, that the X-ray emission of
NGC 3923 is significantly contaminated by foreground/background point
sources. Within a $10\arcmin$ radius of the galaxy center at least 8
point sources are easily identified by visual inspection of the PSPC
image. (Only $r\la 3\arcmin$ is shown in the Figure.) We list the
positions of these sources in Table \ref{tab.src}.

\begin{table}
\caption{Identified Point Sources}
\label{tab.src}
\begin{tabular}{ccc}
Source & R. A. & Decl.\\ 
1 & 11$^{\rm h}$ 51$^{\rm m}$ 06$^{\rm s}$ & $-28\degr$ $46\arcmin$
$47\arcsec$\\
2 & 11 50 58 & -28 43 58\\
3 & 11 50 15 & -28 53 22\\ 
4 & 11 51 04 & -28 57 30\\
5 & 11 51 29 & -28 43 51\\
6 & 11 51 33 & -28 47 41\\
7 & 11 51 35 & -28 56 23\\
8 & 11 51 40 & -28 46 24\\
NWa & 11 50 53 & -28 47 04\\
NWb & 11 50 58 & -28 46 30\\
\end{tabular}

\medskip

These sources (expressed in J2000 coordinates) were identified from
visual inspection of the PSPC image within a $10\arcmin$ radius of the
galaxy center. They were removed from the image for spatial analysis
as described in \S \ref{pt}.

\end{table}

For our analysis of the isophote ellipticities and orientations (\S
\ref{e0}) and the radial profile (\S \ref{radpro}) of the X-ray
surface brightness, we wish to analyze only the distribution of the
diffuse emission associated with NGC 3923 and thus these contaminating
sources must be removed. The most important contaminating source is
number (1) (which lies along the optical major axis to the N-E) since
it lies closest to the galactic center where the S/N , and thus
constraints on the surface brightness, are best; this source is also
apparent in the HRI image. Next in importance is the extended emission
approximately $2.5\arcmin$ to the N-W along the optical minor axis
which appears to consist of emission from at least two point
sources. These candidate sources are listed as NWa,b in Table
\ref{tab.src}.

Our preferred method for removing sources, which is well suited for
analyzing quadrupole moments of X-ray images \cite{bt96}, is to first
choose an annulus around each source to estimate the local
background. Then a second order polynomial surface is fitted to the
background which then replaces the source. We removed sources 2-8
using this method, but sources (1) and NWa,b require some
elaboration. 

Since source (1) affects ellipticity measurements for $r\ga 100\arcsec$
we examined how robust the ellipticity and position angle were to the
method used to remove the source. Another method to remove sources is
by ``symmetric substitution'' (see Strimple \& Binney 1979; Buote \&
Tsai 1995; Buote \& Canizares 1996a). This method exploits the assumed
symmetry of the hot gas distribution. If the gas is approximately
ellipsoidal, then we can replace source (1) with the corresponding
emission obtained by reflecting the source over the galactic center;
i.e. essentially on the other side of the major axis $(a\rightarrow
-a)$. Fortunately, we find that the ellipticities and position angles
are virtually the same whether we remove source (1) by subtracting a
model for the local background or by symmetric substitution. (We use
the former method for ensuing analysis.)

The emission associated with NWa,b cannot be so easily
removed because it is extended, and it is not obvious how to define a
background model. Fortunately this emission only begins to affect the
ellipticities for semi-major axes $\ga 120\arcsec$. At these distance
the S/N does not allow accurate constraints. However, this emission
does need to be removed from the radial profile, and thus we iterate
the local background method to remove the emission associated with
NWa,b. This method is suitable for the radial profile since the
azimuthal averaging is not overly sensitive to small non-axisymmetric
residuals. (We mention the effect of removing this emission on the
radial profile in \S \ref{radpro}.)

\section{X-ray isophote shapes and orientations}
\label{e0}

\begin{table*}
\begin{minipage}{105mm}
\caption{PSPC Ellipticities and Position Angles}
\label{tab.pspc}
\begin{tabular}{rrrllrllr}
\multicolumn{1}{c}{$a$} \\
\multicolumn{1}{c}{(arcsec)} && \multicolumn{1}{c}{$\epsilon_M$} &
\multicolumn{1}{c}{68\%} & \multicolumn{1}{c}{90\%} &
\multicolumn{1}{c}{$\theta_M$} & \multicolumn{1}{c}{68\%} &
\multicolumn{1}{c}{90\%} & \multicolumn{1}{c}{Counts}\\ 
30$\ldots\ldots$  &&  0.10 & 0.04-0.19 & 0.03-0.22 & 42 & 19-55 &  11-77  & 1116\\
40$\ldots\ldots$  &&  0.12 & 0.09-0.15 & 0.07-0.17 & 25 & 15-41 &  07-53  & 1331\\
50$\ldots\ldots$  &&  0.07 & 0.05-0.12 & 0.03-0.13 & 31 & 20-52 &  09-66  & 1460\\
60$\ldots\ldots$  &&  0.08 & 0.05-0.15 & 0.03-0.19 & 57 & 38-81 &  26-101 & 1564\\
75$\ldots\ldots$  &&  0.15 & 0.12-0.19 & 0.09-0.21 & 51 & 42-58 &  36-65  & 1670\\
90$\ldots\ldots$  &&  0.15 & 0.11-0.20 & 0.06-0.23 & 48 & 37-60 &  31-71  & 1754\\
110$\ldots\ldots$ &&  0.10 & 0.08-0.16 & 0.05-0.19 & 39 & 26-61 &  13-156 & 1870\\
\end{tabular}

\medskip

The values of $\epsilon_M$ (and confidence limits) are computed within
an aperture of semi-major axis $a$ on the image with the background
included; the counts, however, have the background subtracted. The
values of $\theta_M$ are given in degrees N through E.

\end{minipage}
\end{table*}

\begin{table*}
\begin{minipage}{105mm}
\caption{HRI Ellipticities and Position Angles}
\label{tab.hri}
\begin{tabular}{rrrllrllr}
\multicolumn{1}{c}{$a$} \\
\multicolumn{1}{c}{(arcsec)} && \multicolumn{1}{c}{$\epsilon_M$} &
\multicolumn{1}{c}{68\%} & \multicolumn{1}{c}{90\%} &
\multicolumn{1}{c}{$\theta_M$} & \multicolumn{1}{c}{68\%} &
\multicolumn{1}{c}{90\%} & \multicolumn{1}{c}{Counts}\\ 
12$\ldots\ldots$  && 0.25 & 0.13-0.32 & 0.08-0.39 & 69 & 44-83  & 30-92  & 309\\  
17$\ldots\ldots$  && 0.26 & 0.14-0.31 & 0.08-0.36 & 59 & 50-76  & 43-88  & 412\\ 
23$\ldots\ldots$  && 0.22 & 0.09-0.27 & 0.05-0.32 & 66 & 47-81  & 34-102 & 506\\ 
32$\ldots\ldots$  && 0.13 & 0.07-0.21 & 0.04-0.26 & 63 & 35-82  & 15-107 & 631\\ 
44$\ldots\ldots$  && 0.14 & 0.11-0.27 & 0.07-0.31 & 76 & 41-87  & 31-103 & 735\\ 
60$\ldots\ldots$  && 0.16 & 0.10-0.28 & 0.05-0.33 & 74 & 44-101 & 33-123 & 825\\ 
\end{tabular}

\medskip

See Table \ref{tab.pspc}.

\end{minipage}
\end{table*}

\begin{figure*}
\parbox{0.49\textwidth}{
\centerline{\psfig{figure=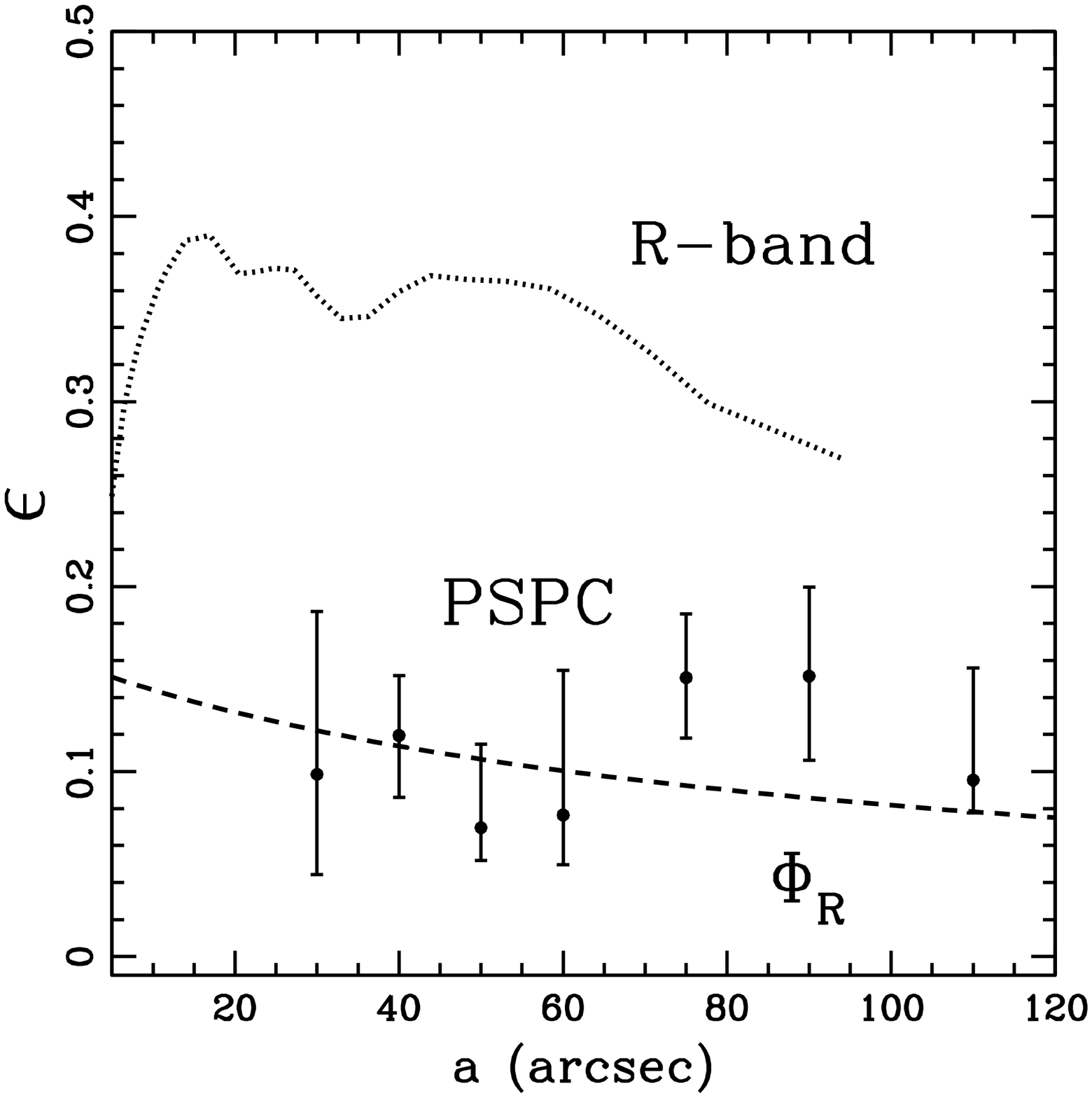,angle=0,height=0.3\textheight}}
}
\parbox{0.49\textwidth}{
\centerline{\psfig{figure=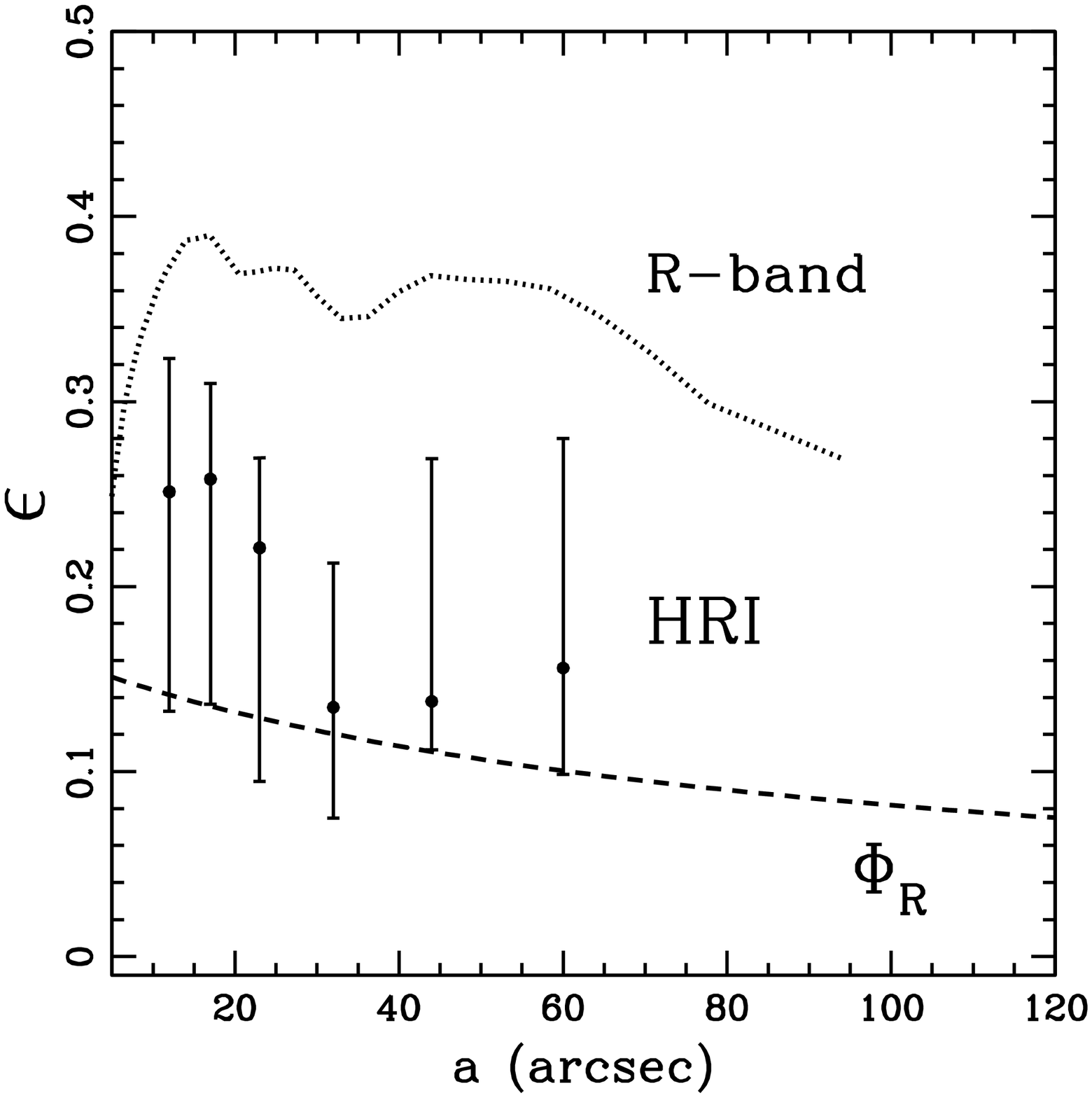,angle=0,height=0.3\textheight}}
}
\caption{\label{fig.e0} Moment ellipticities and $1\sigma$ errors for
PSPC (left) and HRI (right) data.  The isophotal ellipticities of the
$R$-band light from Jedrzejewski (1987) are indicated by the dotted
line. The dashed line represents the isopotential ellipticities of,
$\Phi_R$, the potential assuming $M\propto L_R$ (see \S \ref{gt}).}
\end{figure*}

As is typical for current X-ray data of early-type galaxies the small
number of counts $(\sim 1000)$ for the PSPC and HRI images of NGC 3923
implies that we can only hope to measure with any precision the
ellipticity and position angle of the aggregate X-ray surface
brightness in a large aperture. The method we employ is an iterative
procedure \cite{cm} and is analogous to computing the two-dimensional
moments of inertia within an elliptical region where the ellipticity,
$\epsilon_M$, is given by the square root of the ratio of the
principal moments and the orientation of the principal moments gives
the position angle, $\theta_M$ (see Buote \& Canizares 1994 for
application to {\sl ROSAT} images). The parameters $\epsilon_M$ and
$\theta_M$ are good estimates of the ellipticity $(\epsilon)$ and
position angle ($\theta$, or P.A.) of an intrinsic elliptical
distribution of constant shape and orientation. For a more complex
distribution $\epsilon_M$ and $\theta_M$ are average values weighted
heavily by the outer parts of the regions.

We estimate the uncertainties on $\epsilon_M$ and $\theta_M$ using a
Monte Carlo procedure described in Buote \& Canizares
\shortcite{bc96b}. In sum, the procedure involves constructing 1000
realizations of the PSPC and HRI images taking into account
statistical noise and unresolved sources; these unresolved sources
are modeled according to the $\log N(>S)-\log S$ distribution given by
Hasinger et al. \shortcite{hasinger} and their profiles are given by
the appropriate PSPC or HRI PSF. The 90\% confidence limits on
$\epsilon_M$, for example, are defined by the 5th and 95th percentile
values computed from the 1000 simulations.

The profiles of $\epsilon_M$ and $\theta_M$ are listed for the PSPC in
Table \ref{tab.pspc} and for the HRI in Table \ref{tab.hri}. We also
display the $\epsilon_M$ profiles and their 68\% uncertainties in
Figure \ref{fig.e0}. The apertures listed in the tables are chosen so
that each increment in semi-major axis $a$ consists of approximately
100 source counts. We restrict the PSPC profile to $a\le 110\arcsec$
because of contamination from sources NWa,b (see \S
\ref{pt})\footnote{We emphasize that the bright point source (1) (see
Table \ref{tab.src} and Section \ref{pt}) does not affect the
ellipticity measurements for $a\le 90\arcsec$; i.e. we also measure
$\epsilon_M=0.15$ for $a=75\arcsec,90\arcsec$ on the image including
source (1). However, the values at larger radii differ considerably
depending on whether the source is removed or not; e.g.  at
$a=120\arcsec$, we obtain $\epsilon_M=0.3$ with the source and
$\epsilon_M=0.1$ without the source. As a result, when comparing model
ellipticities to data in future sections of this paper we use only
the ellipticities for $a\le 90\arcsec$ which are robust to the
inclusion or exclusion of source (1).}.

The values of $\epsilon_M$ for the HRI have considerably larger
estimated uncertainties than for the PSPC. In their regions of
overlap, the HRI and PSPC give consistent ellipticities. For the
smallest $a\la 20\arcsec$, the HRI gives some indication of large
ellipticity, $\epsilon_M \ga 0.20$, but the 90\% lower limits are less
than 0.10; at these radii residual aspect errors could be
important. At these $a$ the PSPC is blurred by the PSF, but at $a\sim
70\arcsec-90\arcsec$ ($\sim 10$ kpc) the PSPC gives the best
constrained ellipticity for either the HRI or PSPC of
$\epsilon_M\approx 0.15\pm 0.05$. Although significant, this
$\epsilon_M$ is less than that measured on similar scales for NGC 720
($\sim 0.25$; Buote \& Canizares 1994; 1996b) and for NGC 1332 ($\sim
0.20$; Buote \& Canizares 1996a). 

Like NGC 1332, the P.A. profiles of the PSPC and HRI data are
consistent with the optical value of $\sim 49\degr$ \cite{jed}, albeit
within the rather large uncertainties. Moreover, the PSPC P.A. for
$a\ga 90\arcsec$ may be affected by the bright source (1) and NWa,b
(see Table \ref{tab.src}), and thus it would be premature to rule out
a P.A. twist like that in NGC 720 until higher resolution, better S/N
data (e.g.  {\sl AXAF}) is obtained.

Finally, over the region $a\le 110\arcsec$ we find no obvious
asymmetries in the (PSPC or HRI) surface brightness
(Figure \ref{fig.n3923}) suggestive of unresolved sources or
environmental effects like ram pressure or tidal distortions; e.g.
the centroids of the apertures are very steady over the semi-major
axis listed in Tables \ref{tab.pspc} and \ref{tab.hri}. We also have
divided up the region $r \le 75\arcsec$ into quadrants (aligned with
the major axis of NGC 3923) and have found the counts in each quadrant
to be consistent with each other within $(1-2)\sigma$ errors, similar
to the agreement seen for NGC 720 (Table 2 of Buote \& Canizares
1996b).

\section{Radial profile of X-ray surface brightness}
\label{radpro}

In order to construct the azimuthally averaged radial profiles for the
PSPC and HRI data we require measurements of their respective
background levels. We selected annular regions centered on the galaxy
that are sufficiently far from the galaxy center so that contamination
from the galaxy is minimal. (Also, any point sources were removed.)
We obtain a background rate of $2.4\times 10^{-4}$ cnts s$^{-1}$
arcmin$^{-2}$ for the PSPC and a rate of $3.3\times 10^{-3}$ cnts
s$^{-1}$ arcmin$^{-2}$ for the HRI.

\begin{figure*}
\parbox{0.49\textwidth}{
\centerline{\psfig{figure=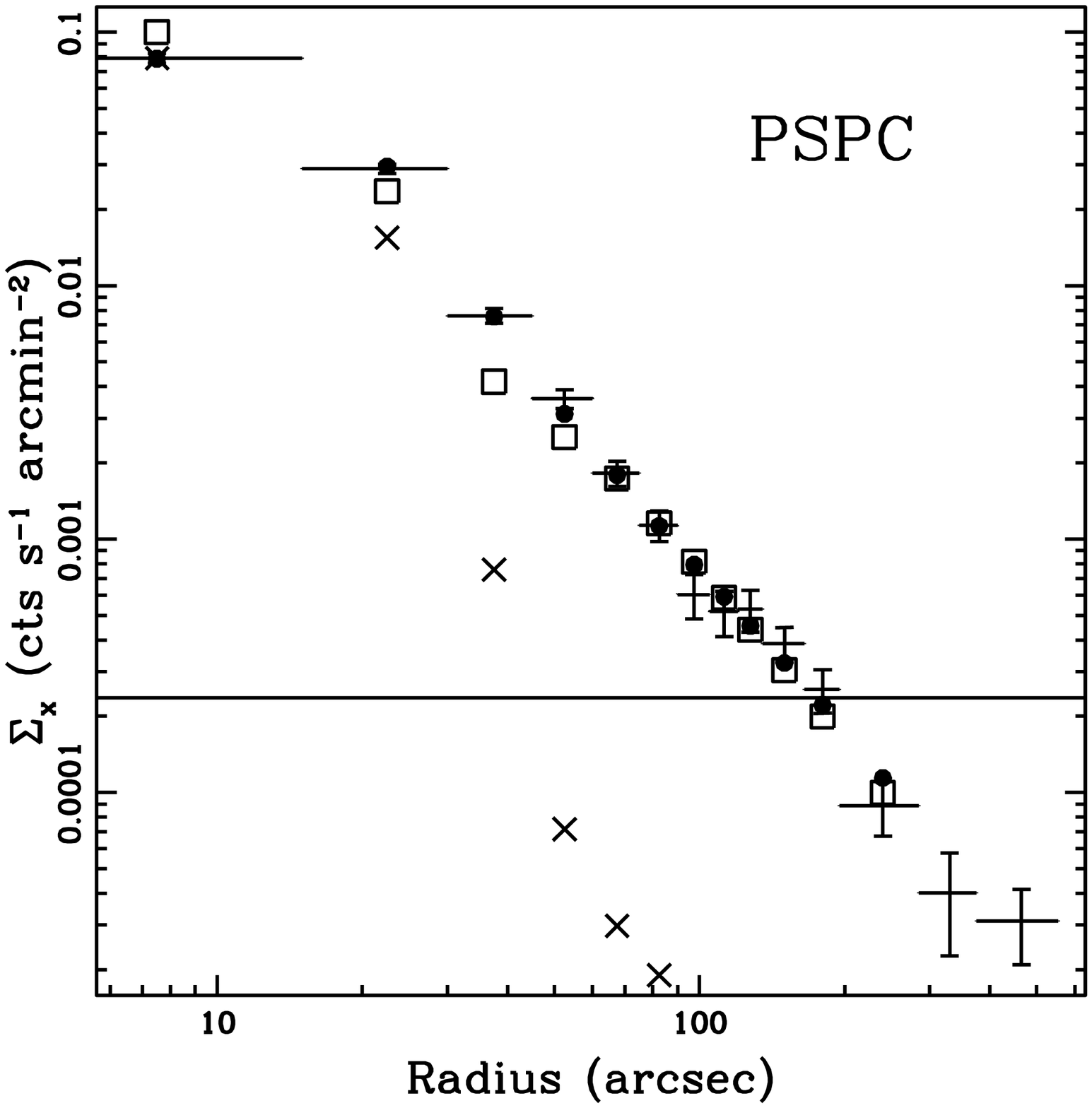,angle=0,height=0.3\textheight}}
}
\parbox{0.49\textwidth}{
\centerline{\psfig{figure=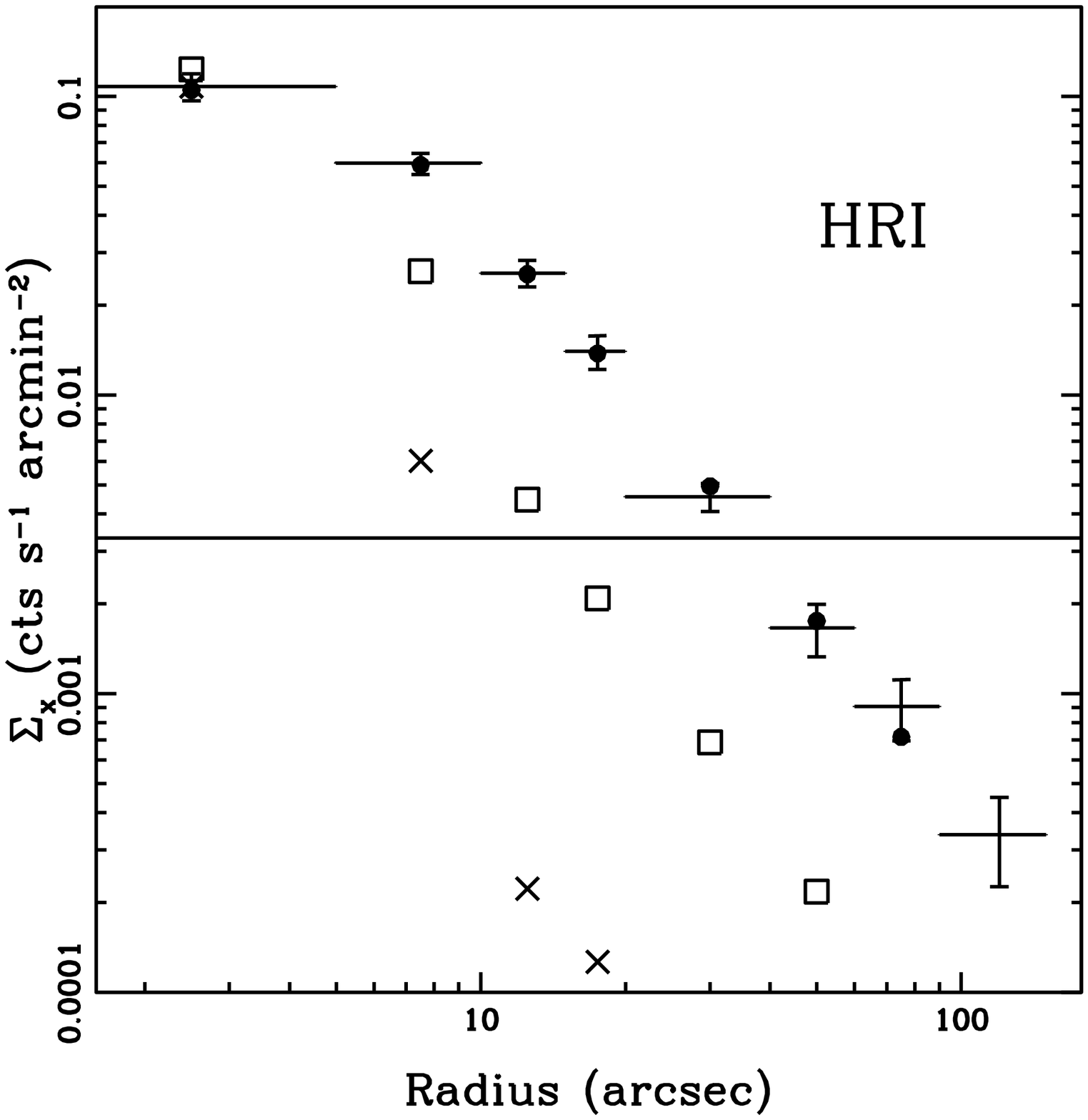,angle=0,height=0.3\textheight}}
}
\caption{\label{fig.radpro} Radial surface brightness profiles of the
PSPC (left) and HRI (right) data. The error bars indicate the data and
the horizontal lines through each point show the bin sizes. The solid
horizontal lines across each plot indicate the background level. The
best-fit $\beta$ model (filled circles), the PSFs (crosses), and the
$R$-band light convolved with the respective PSFs (boxes) are also
shown. All of these quantities have been binned as the X-ray data.}
\end{figure*}

We binned the radial profiles so that each bin had approximately the
same S/N. However, for the innermost bins the S/N is larger because we
did not oversample the respective PSFs. The centers of the radial bins
for both the PSPC and HRI data were determined by the centroid of the
circular region containing $\sim 80\%$ of the total flux; in neither
case did this choice critically affect the radial profile shape. We
display the radial profiles in Figure \ref{fig.radpro} along the the
PSFs; note that residuals of the emission near sources NWa,b may be
seen in the PSPC profile for $r\sim 130\arcsec - 150\arcsec$. The
X-ray emission is clearly extended for both data sets.

A convenient parametrisation of the X-ray radial profiles of
early-type galaxies is given by the ``$\beta$ model'',
$\Sigma_x\propto (R_c^2+R^2)^{-3\beta+0.5}$ \cite{beta}. The model
assumes that the gas is isothermal and that the stars, considered as
test particles, follow a King law. Although these assumptions are
unlikely to be strictly valid for ellipticals, this model provides a
reasonable description of the {\it ROSAT} radial profiles of both NGC
720 \cite{bc96b} and NGC 1332 \cite{bc96a} as well as many other
galaxies (e.g.  Forman et al. 1985). Moreover, for {\sl ROSAT} data
the $\beta$ model typically gives fits of quality very similar to more
sophisticated mass models and thus it serves as a good benchmark for
these more general models (as in \S \ref{models}).

\renewcommand{\arraystretch}{1.3}

\begin{table}
\caption{Simple $\beta$ Model Fits to Radial Profile}
\label{tab.radpro}
\begin{tabular}{lccccc}
& $R_c$\\
Data & (arcsec) & $\beta$ & $\chi^2$ & dof & $\chi^2_{\rm red}$\\ 
PSPC & $5.6_{-1.0}^{+1.1}$ & $0.47_{-0.02}^{+0.02}$ &  8.9 & 9 & 1.0\\
HRI & $5.1_{-1.2}^{+1.6}$ & $0.46_{-0.04}^{+0.05}$ &  1.4 & 4 & 0.4\\
PSPC + HRI & $5.4_{-0.8}^{+0.8}$ & $0.46_{-0.01}^{+0.02}$ & 10.7 & 15 & 0.7\\ 
\end{tabular}
\medskip

The best-fit values and 90\% confidence limits on one interesting
parameter are listed for $R_c$ and $\beta$. The PSFs of each data set
have been incorporated into the fits.

\end{table}

\renewcommand{\arraystretch}{1.0}

\begin{figure*}
\centerline{\psfig{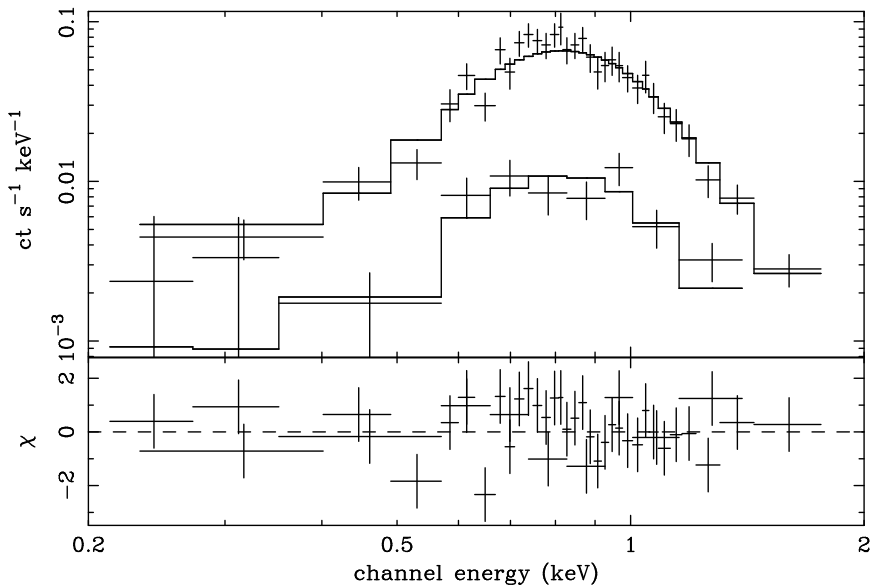}}
\caption{\label{fig.spec} Absorbed single-temperature MEKAL model fit
jointly to the inner (upper curve) and outer (lower curve) regions
(see \S \ref{spectra}). Only the June, 1993 data is shown.}
\end{figure*}

In Table \ref{tab.radpro} we list the results of fitting the $\beta$
model (convolved with the appropriate PSF) to the X-ray data; only
those bins with $S/N\ga 5$ were included in the fits.  When fitted
separately to the PSPC and HRI data, the $\beta$ model gives a good
quality fit and the derived $R_c$ and $\beta$ parameters are similar;
note that if the emission from sources NWa,b is not removed from the
PSPC data, the parameters of these fits are essentially unaffected but
the value of $\chi^2$ increases to $17.4$.

The best-fit model obtained from jointly fitting the PSPC and HRI data
is shown in Figure \ref{fig.radpro}.  The $\beta$ model provides a
reasonably good fit over a large radius range for both data sets.  For
comparison, we plot the $R$-band data from Jedrzejewski
\shortcite{jed} in Figure \ref{fig.radpro}. We have binned the optical
data like the X-rays and have convolved with the appropriate PSFs; for
radii larger than the limiting $a=103\arcsec$ of Jedrzejewski, we
extrapolate the data using the best-fitting De Vaucouleurs model.  For
radii larger than the PSFs, the shape of the optical profile,
$\Sigma_R$, is very similar to the shape of $\Sigma_x$ in agreement
with expectations from steady-state cooling flow models (e.g.  Sarazin
1987). However, the X-ray and optical profiles disagree markedly for
small radii $(r\la 10\arcsec)$ with $\Sigma_R$ being much more
centrally peaked than $\Sigma_x$. Because of its smaller PSF this
effect is more pronounced in the HRI data.

In particular, it is clear that a population of discrete X-ray sources
distributed like $\Sigma_R$ cannot contribute significantly to the
emission. If we assume that the hot gas is described by a $\beta$
model or related model (see \S \ref{models}) then adding such a
discrete component only worsens joint fits to the PSPC and HRI
data. We find that $f_{hg}/f_{disc} > 3.6$ (90\% confidence), where
$f_{hg}$ is the flux of the hot-gas component ($\beta$ model) and
$f_{disc}$ is the flux of the discrete sources where the 0.4-2 keV
flux is computed within a circle of $r=2\arcmin$.

Spectral analysis of {\sl ASCA} data of NGC 3923 by Buote \& Fabian
\shortcite{bf} shows that two temperature models are required by the
data with a cold component, $T_{\rm C}=0.55$ keV, and a hot component,
$T_{\rm H}=4.2(>2.2)$ keV (90\% confidence); the ratio of the 0.5-2
keV flux of the cold component to the hot component is 1.9 (1.3-2.8)
at 90\% confidence. If the discrete sources are indeed distributed
like $\Sigma_R$, then the emission of the hot component cannot be
entirely due to discrete sources. This would indicate that the
emission of the hot component is actually largely due to another phase
of the hot gas and that $T_{\rm H}$ is overestimated due to an
artifact of fitting low S/N data as suggested by Buote \&
Fabian. (Note that the cold-to-hot flux ratio derived by Buote \&
Fabian in the 0.5-2 keV band remains essentially the same when
computed in the 0.4-2 keV band analyzed in our paper.)

\section{Spectral analysis}
\label{spectra}

Spectral analysis of the X-ray data is required for our study of the
mass distribution in NGC 3923 to determine (1) how much of the
emission is due to hot gas, and (2) the temperature profile of the
gas. The superior spectral resolution of {\sl ASCA} is better suited
than {\sl ROSAT} to address issue (1), and as discussed at the end of
the previous section, the {\sl ASCA} spectrum is consistent with $\sim
35\%$ of the X-ray emission in the 0.5-2 keV band arising from a
population of discrete sources \cite{bf}. Since this much discrete
emission is inconsistent with the {\sl ROSAT} radial profiles we
shall neglect it and assume all of the emission arises from hot
gas. This assumption does not seriously affect analysis of the mass
distribution if the discrete contribution is $\la 20\%$ \cite{bc97a}.

Unlike the {\sl ASCA} data, the {\sl ROSAT} PSPC data allows us to
directly measure any temperature gradients. To investigate this issue
we analyzed the PSPC spectra in a circular region with $r=30\arcsec$
and an annular region with $r=60\arcsec-120\arcsec$; the spectra of
these regions are shown in Figure \ref{fig.spec}. We fit a model
consisting of Galactic absorption $(N_H=6.4\times 10^{20}$ cm$^{-2}$
-- Stark et al. 1992) and a thin thermal plasma. Our thermal plasma
model is the MEKAL model which is a modification of the original MEKA
code (Mewe, Gronenschild, \& van den Oord 1985; Kaastra \& Mewe 1993)
where the Fe-L shell transitions crucial to the X-ray emission of
ellipticals have been re-calculated \cite{mekal}. We take solar
abundances according to Anders \& Grevesse \shortcite{ag} and
photo-electric absorption cross sections according to
Baluci\'{n}ska-Church \& McCammon \shortcite{phabs}.

All of the spectral fitting was implemented with the software package
XSPEC \cite{xspec} using the $\chi^2$ minimization method. In order
for the weights to be valid for the $\chi^2$ method we regrouped the
PI bins such that each group contained at least 20 counts. We
restricted the fits to energies above 0.2 keV because of the low S/N
coupled with uncertainties in the PSPC response at these lowest
energies. We fitted the December and June observations jointly, but
with their normalizations as free parameters.

\renewcommand{\arraystretch}{1.35}

\begin{table}
\caption{Spectral Fits}
\label{tab.spec}
\begin{tabular}{lccrrr}
Name & $T$ & $Z$ & $\chi^2$ & dof & $\chi^2_{\rm red}$\\
     & (keV) & $(Z_{\sun})$\\
$0\arcsec - 30\arcsec$ & $0.50_{-0.04}^{+0.05}$ &
     $0.44_{-0.15}^{+2.3}$ & 51.3 & 50 & 1.0\\ 
$60\arcsec - 120\arcsec$ & $0.53_{-0.15}^{+0.17}$ &
     $0.09_{-0.07}^{+0.67}$ & 7.9 & 12 & 0.6\\ 
BOTH & $0.50_{-0.05}^{+0.04}$ &
     $0.45_{-0.16}^{+1.2}$ & 61.6 & 64 & 1.0\\
\end{tabular}
\medskip

Results of fitting a MEKAL model modified by Galactic photo-electric
absorption. We list 90\% confidence limits on one interesting
parameter.
\end{table}

The results of the spectral fits are given in Table
\ref{tab.spec}. The model is an excellent fit to the low resolution
PSPC data in both regions with the temperatures and abundances giving
consistent values within the uncertainties. There is some indication
that the inner region prefers higher abundances than the outer
region. This may suggest a real abundance gradient, or it may be due
to a difference in S/N \cite{bf}. In any event, a joint fit to the
inner and outer region is formally acceptable and gives parameters
similar to the inner region. Hence, we find no evidence of temperature
gradients, and since the PSPC spectral constraints are even tighter
than found for NGC 720 any gradients consistent with this data are
unimportant for analysis of the mass distributions (see Buote \&
Canizares 1994).

Note that we did jointly fit the PSPC and {\sl ASCA} data and found
that the constraints did not change appreciably from the {\sl ASCA}
results alone, even for two-temperature models.

\section{Geometric Test for dark matter}
\label{gt}

\begin{figure*}
\parbox{0.49\textwidth}{
\centerline{\psfig{figure=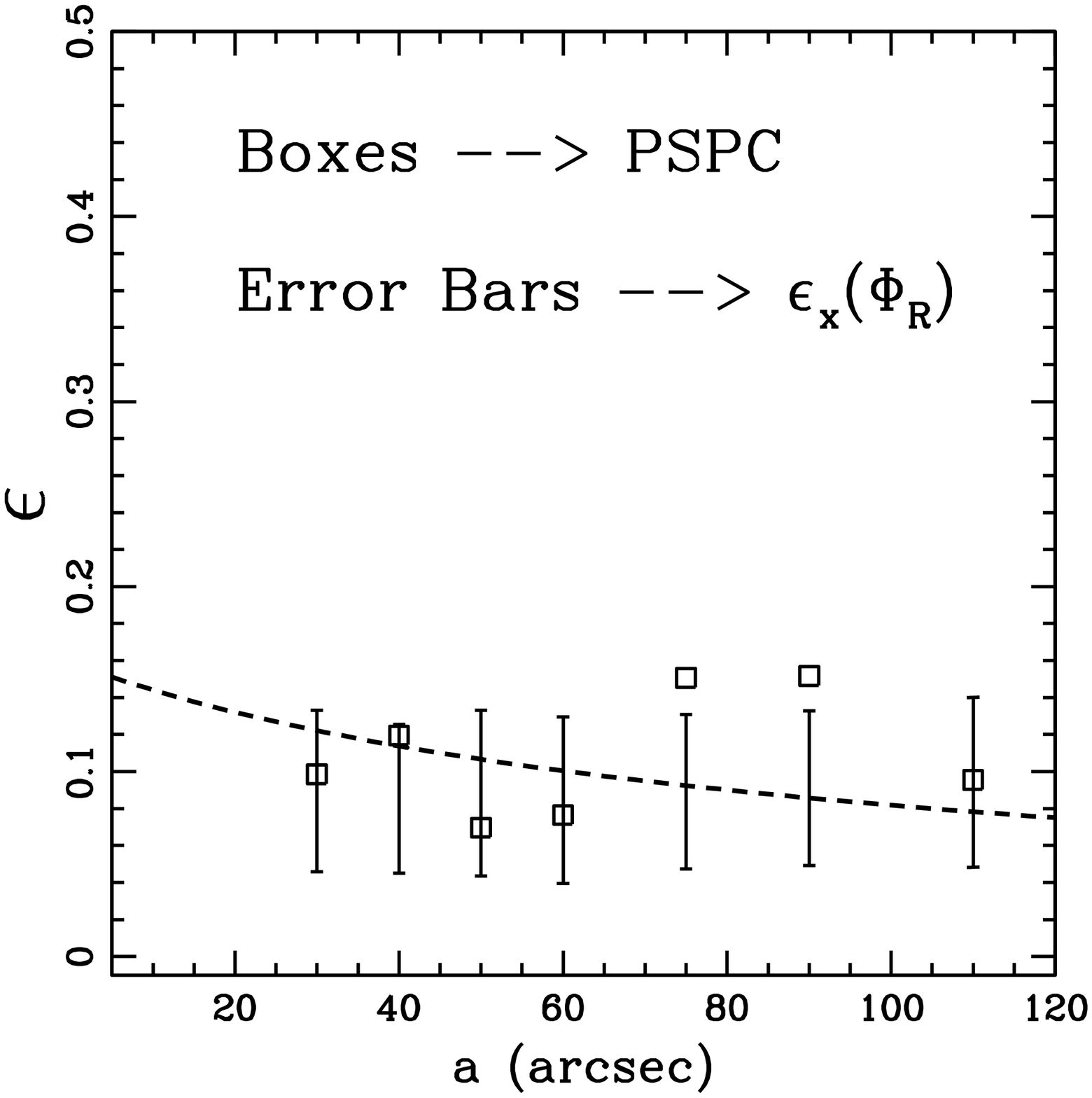,angle=0,height=0.3\textheight}}
}
\parbox{0.49\textwidth}{
\centerline{\psfig{figure=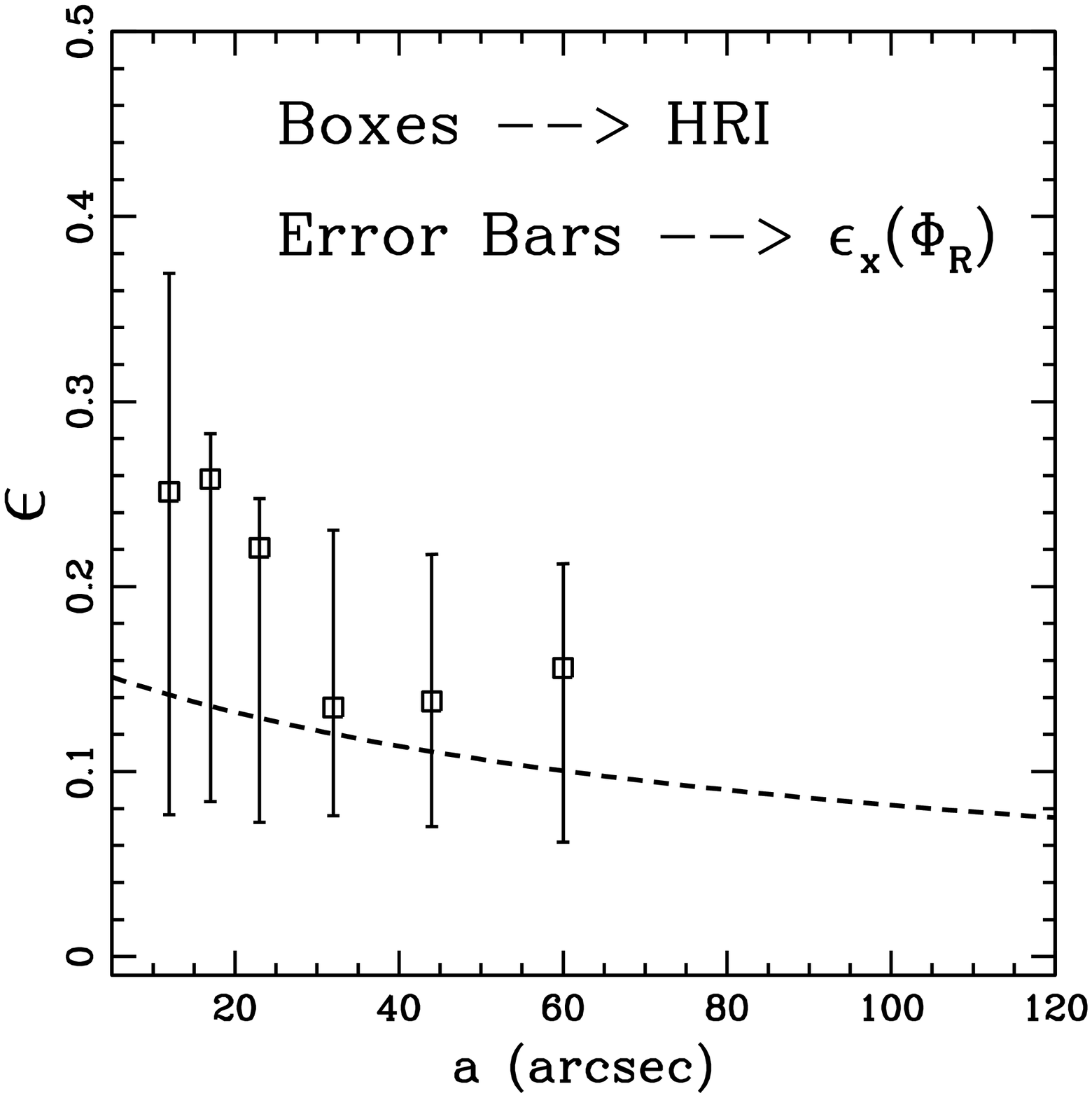,angle=0,height=0.3\textheight}}
}
\caption{\label{fig.gt} Geometric Test for Dark Matter -- see \S \ref{gt}.}
\end{figure*}

The shapes of the X-ray isophotes allow the shape and radial
distribution of gravitating matter to be probed in a way that is more
robust than the traditional spherical approach (for a review see Buote
\& Canizares 1997b).  Assuming the hot gas is approximately in
hydrostatic equilibrium and that the emission is adequately described
by a single phase, then the volume X-ray emissivity, $j_x$, obeys an
``X-ray Shape Theorem'' (Buote \& Canizares 1994, \S 3.1; 1996a, \S
5.1) which states that $j_x$ and the gravitational potential, $\Phi$,
have the same three-dimensional shapes independent of the temperature
profile of the gas. One may thus make a ``Geometric Test'' for dark
matter that is distributed differently from $L$ (with $L$ the
luminosity distribution of the optical stellar light) by comparing the
shape of $\Phi_L$ generated assuming $M\propto L$ with the shape of
$j_x$ obtained from deprojecting the X-ray image data.  As this
comparison is independent of the poorly constrained temperature
profile, $T(r)$, of the gas (see \S \ref{spectra}) this test for dark
matter is more robust than computing the radial mass distribution
which is directly proportional to $T(r)$\footnote{The X-ray shape
analysis is also more robust to the effects of multi-phase gas and
magnetic fields -- see Buote \& Canizares \shortcite{bc97b}.}.

NGC 3923 has negligible stellar rotation, but if rotation of the gas
is dynamically important then we must replace $\Phi$ with the
appropriate effective potential.  Theoretically, even without strong
stellar rotation one may \cite{kley} or may not \cite{nsf} expect a
rotating cooling flow to develop depending on whether angular momentum
of the gas is conserved. Highly flattened X-ray isophotes indicative
of a cooling disk have not been observed in ellipticals and thus we
shall ignore any contribution from gas rotation\footnote{We should
also emphasize that the wiggles displayed by the smoothed X-ray
isophotes in Figure \ref{fig.n3923} are not statistically significant
and thus do not suggest departures of the gas from hydrostatic
equilibrium; i.e. the S/N of the PSPC and HRI data is too poor to
measure deviations of isophote shapes from elliptical.  However, even
if those wiggles in the contours were due to real non-equilibrium
motions in the gas, the hydrostatic assumption is still very likely a
good one. Buote \& Tsai \shortcite{bt95} analyzed the X-ray isophotes
of an X-ray cluster formed in an N-body/hydrodynamical simulation.
They concluded that relatively soon after major mergers the cluster
settled down quickly to a relaxed state which allowed accurate
determination of the shape of the mass from analysis of the X-ray
isophote shapes, even though the simulated clusters had X-ray
isophotes far from elliptical in shape.}.

To apply the Geometric Test we constructed a constant $M/L$ potential
using the $R$-band surface photometry, $\Sigma_R$, of Jedrzejewski
\shortcite{jed}. The major-axis profile of $\Sigma_R$ is well fitted
by a De Vaucouleurs Law with an effective semi-major axis,
$a_e=92\arcsec$. Taking into account the ellipticity of the isophotes,
we fitted the De Vaucouleurs Law to the mean radial profile, where the
mean radius is, $r=\sqrt{ab}=a\sqrt{q}$, where $a$ is the major axis
and $q$ the axial ratio. This gives a mean effective radius,
$R_e=73\arcsec$. Since the 3-D density giving rise to a De Vaucouleurs
Law is well approximated by the Hernquist profile, $\rho\propto
r^{-1}(r_s+r)^{-3}$ \cite{hern}, we consider for our $M\propto L_R$
model a Hernquist profile with
$r_s=R_e/1.8153=40\arcsec$. Furthermore, we take $\rho$ to be
stratified on oblate or prolate spheroids of constant ellipticity,
where we use the intensity weighted ellipticity of $\Sigma_R$,
$\langle \epsilon_R\rangle=0.30$.

This simple prescription for the $M\propto L_R$ model is sufficient
for comparison to the {\sl ROSAT} X-ray data because of the relatively
crude X-ray constraints: i.e. only a global constraint on the isophote
shapes of the PSPC data over a small range in radius, $r\sim
70\arcsec-90\arcsec$, is useful for the comparison -- as we show
below, the HRI data do not provide interesting constraints over the
PSPC data. With better quality data from future missions which
accurately measure ellipticity gradients over a large range in radius
a more accurate $M\propto L_R$ approximation will need to be
considered.

We show in Figure \ref{fig.e0} the ellipticities of the isopotential
surfaces of the potential, $\Phi_R$, assuming $M\propto L_R$ and
edge-on oblate symmetry. These ellipticities are considerably smaller
than those of the $R$-band light because the spherically symmetric
monopole term in the potential dominates the isopotential
ellipticities for $r\ga r_c$
\footnote{See Figure 2-13 in \S 2.3 of Binney \& Tremaine
\shortcite{bin} for a related discussion of isopotential
ellipticities.}. The ellipticities of the X-ray surface brightness,
particularly for the PSPC data for $a\sim 70\arcsec-90\arcsec$, appear
to be significantly larger than the isopotential ellipticities and
thus indicate a failure of the $M\propto L_R$ hypothesis.

However, to rigorously compare $\Phi_R$ to the X-ray data we must
formally deproject $\Sigma_x$ to get the ellipticities of $j_x$. The
procedure we adopt for this comparison, which is suitable for the
relatively crude constraints provided by the X-ray data, is to first
fit a simple model to the radial profile of $\Sigma_x$; for this
purpose, the results of the $\beta$ model fit jointly to the PSPC and
HRI data are suitable (see \S \ref{radpro}). This model for
$\Sigma_x(R)$ is then deprojected to obtain $j_x(r)$\footnote{Note
that the parameters in Table \ref{tab.radpro} correspond to the
deprojected parameters.}. We then assign to $j_x$ the ellipticities of
$\Phi_R$ which gives a spheroidal emissivity distribution,
$j_x(r,\theta)$. This spheroidal $j_x$ is then projected back onto the
sky plane, while adjusting $R_c$ and $\beta$ to maintain a best fit of
the radial profile of $\Sigma_x$ (convolved with the appropriate
instrument PSF), to yield a $M\propto L_R$ model of the X-ray surface
brightness. We compute moment ellipticities, $\epsilon_x(\Phi_R)$, for
1000 Monte Carlo simulations of this model in analogy to the data (see
\S \ref{e0}).

The $1\sigma$ error bars for $\epsilon_x(\Phi_R)$ as well as the
observed X-ray ellipticities and the 3-D isopotential ellipticities of
$\Phi_R$ are displayed in Figure \ref{fig.gt}. Although most of the
measured X-ray ellipticities lie within the error bars, the best
measured PSPC ellipticities from $a\sim 70\arcsec-90\arcsec$ exceed
the values predicted by the edge-on oblate $\Phi_R$ at the 85\%
confidence level. The edge-on prolate $\Phi_R$ is formally discrepant
at the 80\% level at these radii\footnote{We caution that
interpretation of the significance of this discrepancy must take into
account the {\sl PSPC} PSF as we have done; i.e. the two deviant data
points for $a\sim 75\arcsec - 100\arcsec$ are not simply random
fluctuations weighted equally with other points over the radial range
investigated. (Also, care is required in the interpretation of the
last measured $\epsilon_M$ at $a=110\arcsec$ for the PSPC data since
its value depends on the bright point source that is removed -- see
Sections \ref{pt} and \ref{e0}.) This point is illustrated in Figures
13 (a) and (b) of Buote \& Canizares \shortcite{bc96a} for a similar
X-ray analysis of NGC 1332. That is, the X-ray ellipticity profiles
produced by different models are smeared out by the {\sl PSPC} PSF,
and thus a discrepancy is only achieved at radii large enough so that
the ellipticity differences between models exceeds the relatively
large error bars of the measured ellipticity; of course, the radial
range is limited from above by the decreasing S/N.  Thus, the radii
$a\sim 75\arcsec - 100\arcsec$ are where the different models
convolved with the {\sl PSPC} PSF begin to show significant
ellipticity differences and the X-ray data still have good constraints
on the measured ellipticity. NGC 3923, NGC 1332, and NGC 720 are quite
similar in this regard since they have similar length scales and
similar S/N PSPC observations. However, the {\sl HRI} data of NGC 720
are of sufficient quality so that the effect of the PSFs on the
Geometric Test can be clearly seen: see Figure 2 of Buote \& Canizares
\shortcite{bc97b}.}.

Hence, the X-ray isophote shapes indicate that the
$\epsilon_x(\Phi_R)$ model is either too round, too centrally
concentrated, or both; i.e. dark matter is required which is flattened
and probably more extended than $\Sigma_R$.  This discrepancy, though
formally marginal and of lesser significance than found for the other
two early-type galaxies studied NGC 720 and NGC 1332 (e.g.  Buote \&
Canizares 1997a,b), is of precisely the same character: i.e. flattened
and extended dark matter is required in these ellipticals.  Since the
flattest optical isophotes of NGC 3923 are rounder than those of NGC
720 and NGC 1332, it is possible that the symmetry axis of NGC 3923 is
inclined more along the line-of-sight than the other two. This would
account for the marginal significance of the NGC 3923 result if the
intrinsic shapes of the three galaxies are in fact similar. 

We note that, as explained in Section 5.1 of Buote \& Canizares
\shortcite{bc96a}, if an ellipsoidal galaxy is inclined along the line
of sight, then the Geometric Test results give a lower limit to the
true discrepancy. That is, if mass follows light, then $j_x$ and
$\Phi_L$ must be co-axial and thus share the same inclination angle,
$i$. Since the projection of an ellipsoid with $i<90^{\circ}$ is
necessarily rounder than if the galaxy were viewed edge-on,
deprojection of $\Sigma_x$ and $\Sigma_L$ assuming $i=90^{\circ}$ will
yield $j_x$ and $\Phi_L$ which are rounder than in reality. However,
this edge-on deprojection of an inclined ellipsoidal galaxy only means
that any differences in the inferred shapes of $j_x$ and $\Phi_L$ are
smaller than if the true inclination angle were used.

\section{Detailed hydrostatic models}
\label{models}

The results of the Geometric Test from the previous section suggest
the presence of a dark matter halo that is flattened and more extended
than the optical light. To generally constrain the allowed
distribution of gravitating matter, we must employ explicit solutions
of the hydrostatic equation.  As the low S/N of the PSPC and HRI
images does not justify a sophisticated non-parametric inversion of
the data, we instead consider simple intuitive models to place
constraints on the aggregate shape and radial distribution of
gravitating mass.  As in our previous studies (e.g.  Buote \&
Canizares 1997b), we follow the pioneering approach of Binney \&
Strimple \shortcite{bs} and solve the hydrostatic equation assuming a
single-phase, non-rotating, ideal gas,
\begin{equation}
\tilde{\rho}_g=\exp \left[\Gamma\left(1-\tilde{\Phi}\right)\right], \label{eqn.gas} 
\end{equation}
where $\Gamma=\mu m_p\Phi(0)/k_BT$ and where, e.g. 
$\tilde{\rho}_g=\rho_g(\vec x)/\rho_g(0)$. Equation (\ref{eqn.gas})
assumes the gas is isothermal, but Strimple \& Binney \shortcite{sb}
and others (Fabricant, Rybicki, \& Gorenstein 1984; Buote \& Canizares
1994; Buote \& Tsai 1995) have shown that the constraints on the shape
of the gravitating matter distribution are not very sensitive to
temperature gradients. Hence, we shall assume an isothermal gas which
is consistent with the spectral constraints for NGC 3923 (\S
\ref{spectra}) and the temperature profiles of other galaxies (e.g. 
Buote \& Canizares 1994; for a review see Sarazin 1997).

Our procedure begins with a model for the gravitating mass. We
consider spheroidal density distributions whose isodensity surfaces
are concentric, similar spheroids. By examining oblate and prolate
configurations we bracket the intermediate behavior of triaxial
models; i.e. as our important constraints on the models are the X-ray
isophote shapes for semi-major axes $a\sim 75\arcsec$, the detailed
radial behavior of the isophote shapes and orientations which
distinguish triaxial models cannot be usefully constrained by the NGC
3923 PSPC and HRI data.

Once the type of density model and its associated ellipticity,
$\epsilon_{mass}$, are chosen, we compute $\Phi$ and $\rho_g$.  (We
initially focus on one-component models of the density of the
gravitating matter and then consider separate density models for the
dark and luminous matter.)  The model X-ray surface brightness is then
generated by integrating $\rho^2_g$ along the line of
sight\footnote{The plasma emissivity is a very weak function of
temperature when convolved with the spectral response of either the
PSPC or HRI.} assuming the galaxy is viewed edge-on\footnote{We assume
the symmetry axis of the galaxy spheroid lies in the plane of the sky;
i.e. we are not attempting in this analysis to uncover the true
three-dimensional shape, though Binney \& Strimple \shortcite{bs} have
found that the X-ray analysis is not extremely sensitive to small
inclination of the symmetry axis with respect to the sky plane. See
Buote \& Tsai \shortcite{bt95} for a thorough discussion of projection
effects on X-ray shape analysis.}. We model the radial mass as either
a softened isothermal potential with mass density, $\rho\propto (a^2_s
+ a^2)^{-1}$, or a Hernquist profile, $\rho\propto a^{-1}(a_s +
a)^{-3}$, where $a_s$ is the appropriate mass scale length in each
case. These models bracket the interesting range of densities in our
previous X-ray analyses of NGC 720 and NGC 1332\footnote{The Hernquist
density gives fits to the density profiles of halos in Cold Dark
Matter simulations that are very similar to the universal model of
Navarro et al. \shortcite{nfw}.}. We convolve the surface brightness
with the PSF of the appropriate detector and evaluate the model on a
grid of pixels identical to that used for analyzing the data
(\S\ref{obs}). We construct the radial profiles as for the data
(\S\ref{radpro}) and determine $a_s$ and $\Gamma$ by fitting the model
profile jointly to the PSPC and HRI data. (The normalizations of the
PSPC and HRI are free parameters.) By comparing the ellipticity of the
model surface brightness to the data we constrain the input
ellipticity of the gravitating mass.

\begin{table*}
\begin{minipage}{135mm}
\caption{Ellipticity of the Gravitating Matter}
\label{tab.shape}
\begin{tabular}{lccccccccc}

& \multicolumn{2}{c}{Oblate $\epsilon_{mass}$} &
\multicolumn{2}{c}{Prolate $\epsilon_{mass}$} & $a_s$ \\
Model & 68\% & 90\% & 68\% & 90\% & (arcsec) & $|\Gamma|$ & $\chi^2$ &
dof & $\chi^2_{\rm red}$\\
$\rho\sim r^{-2}$ & 0.43-0.59 & 0.35-0.65 & 0.40-0.54 & 0.33-0.58 &
2.2-3.1 & 6.6-7.2 & 16 & 15 & 1.1\\
Hernquist & 0.45-0.61 & 0.36-0.66 & 0.42-0.55 & 0.33-0.59 & 53-75 &
5.7-5.8 & 28 & 15 & 1.9\\ 
\end{tabular}

\medskip

Derived shapes of the gravitating matter for the spheroidal mass
models assuming an isothermal gas (see \S \ref{models}). 90\%
confidence ranges are quoted for the scale length $a_s$ and
$|\Gamma|$. Typical values of $\chi^2$ are listed in the 90\%
intervals. Note for the $M\propto L_R$ model we obtain $\chi^2=56.0$
for 16 dof.

\end{minipage}
\end{table*}

In Table \ref{tab.shape} we give the constraints on the shape of the
gravitating matter obtained from these models. The mass is
significantly flattened, $\epsilon_{mass}\approx 0.5$, with the
derived shapes being essentially the same for both the $\rho\sim
r^{-2}$ and Hernquist models. These ellipticities are inconsistent
with the intensity-weighted ellipticity of the $R$-band light
$(\langle\epsilon_R\rangle=0.30)$ at more than the 90\% level and are
marginally inconsistent with the flattest optical isophotes
$(\langle\epsilon_R^{max}\rangle=0.39)$.  The derived range of values
of $\epsilon_{mass}$ and the fact that they exceed the average
ellipticity of the optical isophotes are consistent with the previous
results we obtained for NGC 720 (and NGC 1332) (e.g.  Buote \&
Canizares 1997a,b).

\begin{figure*}
\parbox{0.49\textwidth}{
\centerline{\psfig{figure=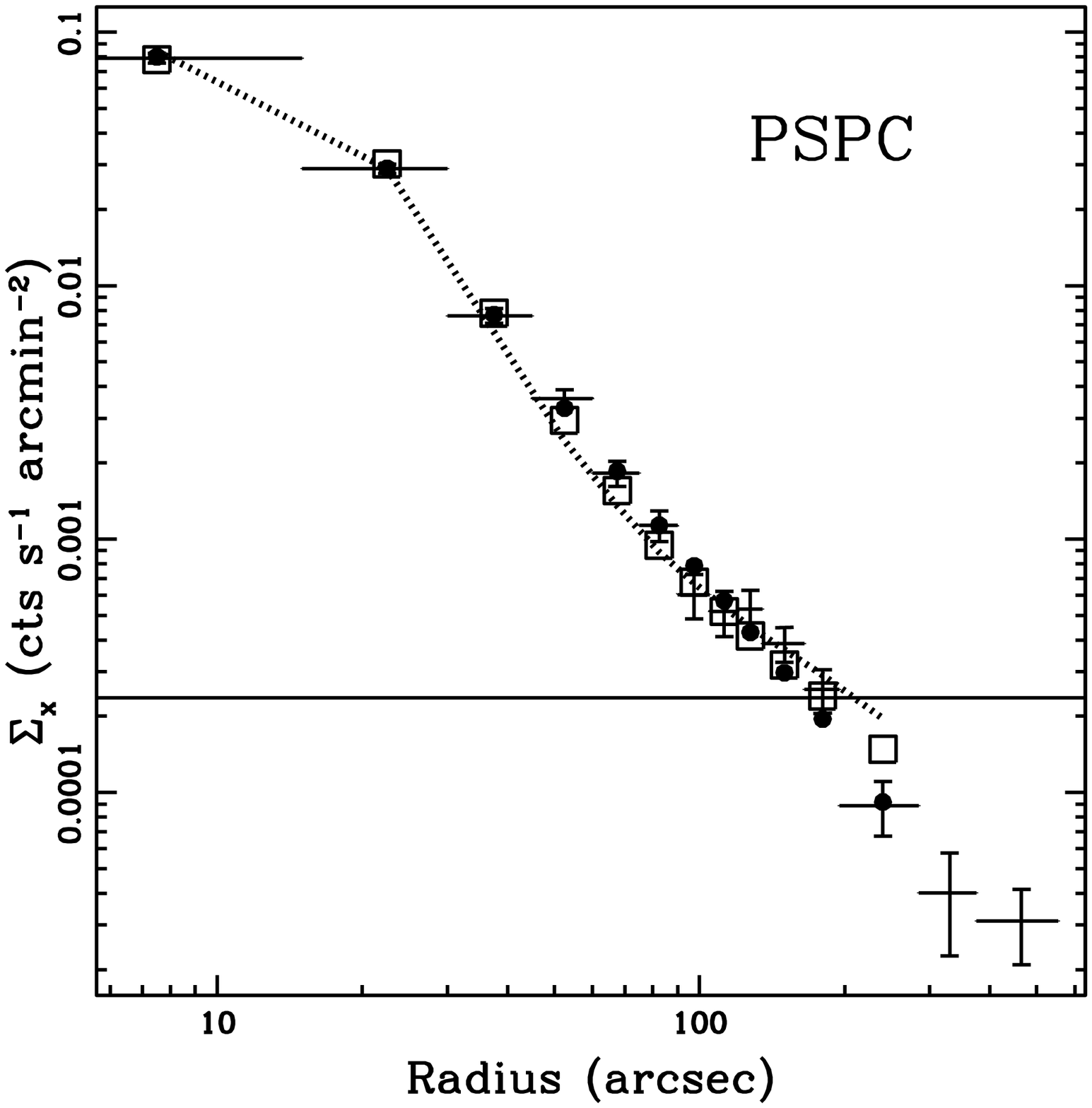,angle=0,height=0.3\textheight}}
}
\parbox{0.49\textwidth}{
\centerline{\psfig{figure=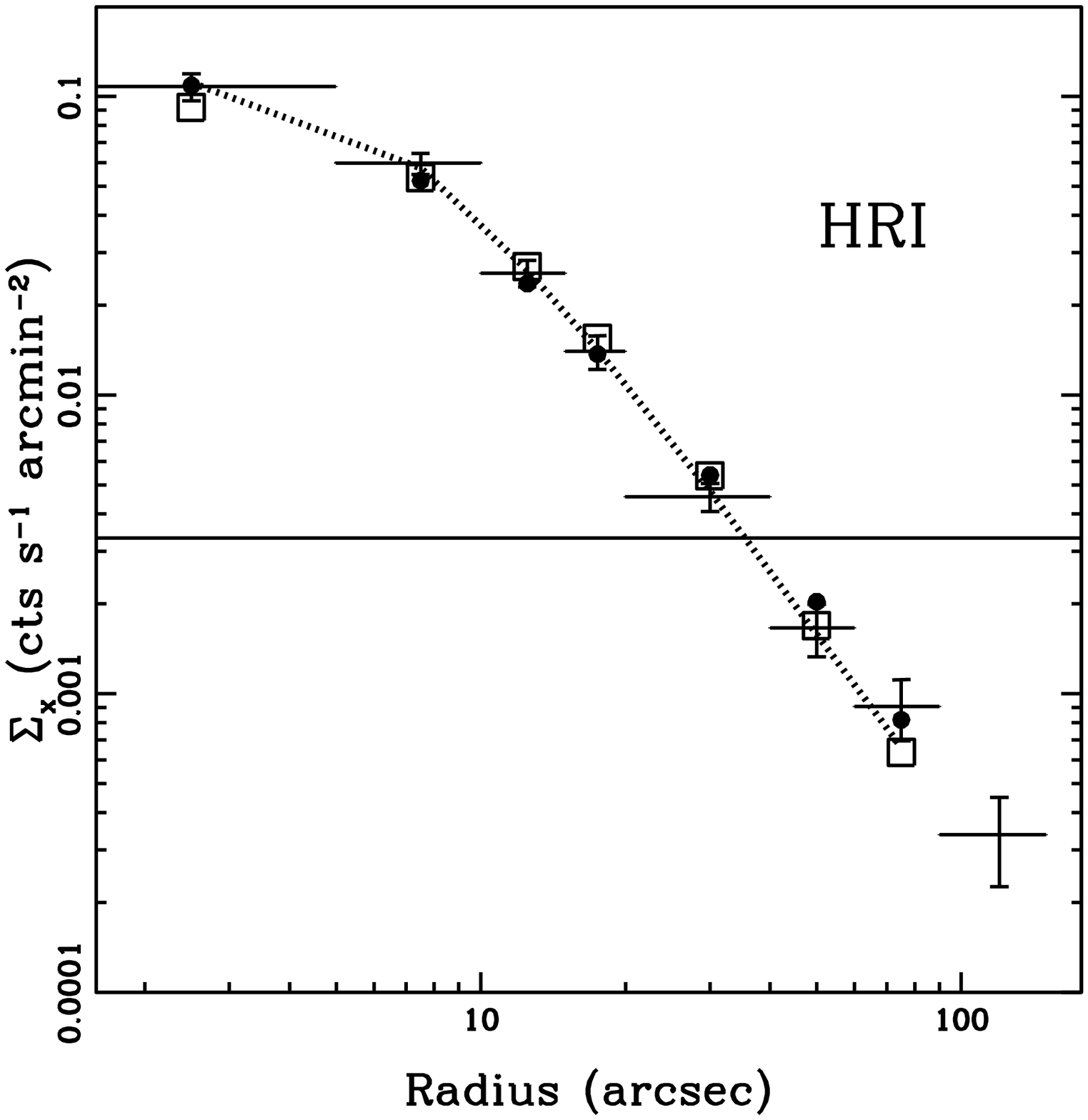,angle=0,height=0.3\textheight}}
}
\caption{\label{fig.radmod} Radial surface brightness profiles of
selected oblate models fit to the PSPC (left) and HRI (right)
data. The error bars indicate the data and the horizontal lines show
the bin sizes.  The filled circles represent the best-fit $\rho\sim
r^{-2}$ model having $\epsilon_{mass}=0.50$; the boxes represent the
best-fit Hernquist model with $\epsilon_{mass}=0.50$; the dotted line
is the best-fitting constant $M/L$ model. The radial profiles of these
models have been binned as the X-ray data.}
\end{figure*}

In Figure \ref{fig.radmod} we plot the radial profiles of typical
$\rho\sim r^{-2}$ and Hernquist models. The $\rho\sim r^{-2}$ model
fits are very similar in quality to those of the $\beta$ model (\S
\ref{radpro}) but with slightly larger values of $\chi^2$. (The
differences are not obvious from visual inspection.) However, the
Hernquist fits are noticeably worse than the $\rho\sim r^{-2}$ model
and have unacceptable $\chi^2_{\rm red}\approx 1.9$. For $r\la 15\arcsec$
the Hernquist model is too flat and then too steep out to $r\sim
100\arcsec$. At larger radii, the small fitted values of $\Gamma\sim
5.8$ force the Hernquist radial profile to be flatter than the data.

These deviations of the Hernquist model are most pronounced for the
$M\propto L_R$ model (see \S \ref{gt}) which has $a_s=40\arcsec$ and a
best-fit $\Gamma=5.5$ and $\chi^2=56.0$ (16 dof). The PSPC
ellipticities of the $M\propto L_R$ model are 0.07 for $a\sim
70\arcsec-90\arcsec$ which is below the 90\% confidence limits of the
data (Table \ref{tab.pspc}). Hence, by assuming the gas is isothermal
we find that the $M\propto L_R$ model is inconsistent with the data at
a significance level greater than indicated by the more robust
Geometric Test (\S \ref{gt}).

It should be emphasized that most of the disagreement between the
$\rho\sim r^{-2}$ and Hernquist models occurs in the innermost
bins. For example, if we exclude the two inner bins for the PSPC
$(r=0\arcsec - 30\arcsec)$ and the three inner bins for the HRI
$(r=0\arcsec - 15\arcsec)$\footnote{We remove a larger region for the
PSPC because of its larger PSF.} we obtain a minimum $\chi^2$ of $\sim
10.0$ for the $\rho\sim r^{-2}$ model and $\chi^2\sim 11.4$ for the
Hernquist model. Since in these innermost regions the surface
brightness profile could be modified by a multi-phase cooling flow or
by magnetic fields, we suggest some caution in interpreting $\chi^2$
values of our models in these central regions until better data can
distinguish between different scenarios.  

However, there is reason for optimism on both of these accounts. The
distortion due to a multi-phase cooling flow is likely to be
insignificant since single-phase analyses generally provide good
descriptions for multi-phase cooling flows in the cores of clusters
(Thomas, Fabian, \& Nulsen 1987; Allen, Fabian, \& Kneib 1996).  Since
no significant radio emission has been detected from NGC 3923 (e.g. 
Birkinshaw \& Davies 1985) it is unlikely that the hot gas is
supported by magnetic fields.  It should also be added that the
$\rho\sim r^{-2}$ model appears to fit the radial profile better than
the Hernquist model for $r\sim 100\arcsec - 300\arcsec$, but the
effect is not highly significant because of the low S/N in this
region\footnote{Note that the limits on $\epsilon_{mass}$ are
essentially unaffected when the central bins are excluded.}.

If we fit the $M\propto L_R$ model and add dark matter following the
$\rho\sim r^{-2}$ model, the fits are marginally improved over the
single-component case with a minimum $\chi^2=13.3$ obtained for a
ratio $M_{DM}/M_R=10$ and $M_{DM}/M_R>3$ (90\% confidence). The
ellipticity of the dark matter, $\epsilon_{DM}$, is essentially that
of $\epsilon_{mass}$ for $M_{DM}/M_R=10$ but increases systematically
so that $\epsilon_{DM}\approx \epsilon_{mass} + 0.10$ at the lower
limit $M_{DM}/M_R=3$.  Here it should be understood that $M_{DM}$
includes only dark matter that is distributed differently from $L_R$
and that the $\chi^2$ values from the radial profile fits for given
ratios $M_{DM}/M_R$ depend on the assumed temperature profile.

The Hernquist profiles are only affected when $M_{DM}/M_R\sim 1$ at
which point the fits are improved to minimum $\chi^2\sim 21$ with
large $a_s\sim 400\arcsec$. The dark matter in these models is
required to be extremely flattened.

Similar to our analysis of the $\beta$ model (\S \ref{radpro}), we
find that adding a discrete component with X-ray emission proportional
to $L_R$ does not improve the fits for the $\rho\sim r^{-2}$ model; we
place a slightly stricter limit of $f_{hg}/f_{disc} > 4.5$ (90\%
confidence). The Hernquist models are actually improved by the
addition of a discrete component such that $\chi^2_{\rm red}=1.2$ for
$f_{hg}/f_{disc}=4.6$. However, essentially all of this improvement
takes place for the central bins discussed above; i.e. if we exclude
those bins from the fits, then the discrete model does not improve the
fits significantly. Hence, a discrete component distributed like $L_R$
does not improve fits to $\Sigma_x$ with the $\rho\sim r^{-2}$ model,
and the improvement observed for the Hernquist model occurs in the
central bins for which other physics (e.g, multi-phase cooling flows,
magnetic fields) may equally affect the models.

\begin{table*}
\begin{minipage}{135mm}
\caption{Gravitating Mass and Gas Mass}
\label{tab.mass}
\begin{tabular}{lccc}
Model & $5h^{-1}_{70}$ kpc & $10h^{-1}_{70}$ kpc & $50h^{-1}_{70}$
kpc\\ 
$\rho\sim r^{-2}$ & $(1.1-2.8)\times 10^{11} h_{70}^{-1}M_{\sun}$ &
$(2.4-6.0)\times 10^{11} h_{70}^{-1}M_{\sun}$ & $(12.2-30.5)\times 10^{11}
h_{70}^{-1}M_{\sun}$ \\ 
Hernquist & $(1.3-3.3)\times 10^{11} h_{70}^{-1}M_{\sun}$ &
$(2.6-7.1)\times 10^{11} h_{70}^{-1}M_{\sun}$  & $(6.0-18.1)\times
10^{11} h_{70}^{-1}M_{\sun}$\\
Gas & $0.14\times 10^9 h_{70}^{-5/2}M_{\sun}$ & $0.45\times 10^9
h_{70}^{-5/2}M_{\sun}$  & $6.4\times 10^9 h_{70}^{-5/2}M_{\sun}$\\ 
\end{tabular}

\medskip

90\% confidence values of the gravitating mass corresponding to the
oblate and prolate models in Table \ref{tab.shape} which include 90\%
uncertainties in the (isothermal) temperature. The statistical errors on the gas
mass are less than 10\%.

\end{minipage}
\end{table*}

In Table \ref{tab.mass} we give the spherically averaged masses
corresponding to the models in Table \ref{tab.shape}. We only quote
masses at a few interesting radii to emphasize that without precise
constraints on the temperature profile (and only an aggregate
constraint on the isophote shapes) we do not really constrain the
detailed mass profile. (Note that the spectral constraints in \S
\ref{spectra} which indicate an approximate isothermal gas apply only
for $r\la 15h_{70}^{-1}$ kpc.)  The masses of the $\rho\sim r^{-2}$
and Hernquist models agree very well within $10h_{70}^{-1}$ kpc as
expected because over much of this region the large scale lengths of
the Hernquist model imply an approximate logarithmic slope of -2.
Assuming the gas to be isothermal out to $r>50h_{70}^{-1}$ kpc, the
$\rho\sim r^{-2}$ model, with its flatter slope, has approximately
twice the mass of the Hernquist model at that distance. The gas mass
is less than 1\% of the gravitating mass over the entire radius range
investigated.

Using the total $B$-band luminosity from Faber et al. \shortcite{7s}
scaled to 30 Mpc, $L_B = 3.45h^{-2}_{70}\times 10^{43}$ erg cm$^{-2}$
s$^{-1} = 6.95h^{-2}_{70}\times 10^{10}L_{\sun}$, we have that the
$B$-band mass-to-light ratio in solar units is $\sim 5
h_{70}M_{\sun}/L_{\sun}$ at $10h_{70}^{-1}$ kpc for the $\rho\sim
r^{-2}$ and Hernquist models. At $r=50h_{70}^{-1}$ kpc, the $\rho\sim
r^{-2}$ model has $M/L_B\sim 32h_{70}M_{\sun}/L_{\sun}$ while the
Hernquist model has $M/L_B \sim 17h_{70}M_{\sun}/L_{\sun}$. These
$M/L_B$ values within $10h_{70}^{-1}$ kpc are very consistent with
stellar dynamical studies of ellipticals (e.g. van der Marel
1991). The larger (extrapolated) $M/L_B$ values at $50h_{70}^{-1}$ kpc
agree with previous X-ray studies of ellipticals (e.g. Sarazin 1997)
indicating that NGC 3923 has a mass distribution typical for a galaxy
of its luminosity.

\section{Conclusions}
\label{conc}

We have analyzed the gravitating matter distribution of the E4 galaxy
NGC 3923 using archival X-ray data from the {\sl ROSAT} PSPC and HRI.
Analysis of the PSPC data, which allows more precise constraints than
the HRI data, demonstrates that the X-ray isophotes are significantly
elongated with ellipticity $\epsilon_x=0.15 (0.09-0.21)$ (90\%
confidence) for semi-major axis $a\sim 10h^{-1}_{70}$ kpc and have
position angles aligned with the optical isophotes within the
estimated uncertainties. A bright point source located $\sim
100\arcsec$ along the major axis inhibits reliable ellipticity
constraints for larger radii.

By applying a ``Geometric Test'' for dark matter, which essentially
compares the shapes of the observed X-ray isophotes to those predicted
if mass traces the optical light $L$ (independent of the poorly
constrained temperature profile of the gas), we found that the
ellipticity of the PSPC X-ray surface brightness exceeds that
predicted by the constant $M/L$ hypothesis at the 80\%-85\% confidence
level. The ``Geometric Test'' result is conservative since it only
considers signatures of dark matter that are distributed differently
from the optical light.

Although the evidence for dark matter from the Geometric Test is
marginal, the results from models which employ an explicit solution of
the hydrostatic equation assuming an isothermal gas (which is
supported by the PSPC spectrum -- \S \ref{spectra}) indicate that
$M\propto L$ is highly inconsistent with the radial profiles of the
PSPC and HRI data ($\chi^2_{\rm red}=3.5$ for 16 dof). This particular
discrepancy arises because $L$ is too centrally concentrated: the
derived scale length of the gravitating matter is approximately 1.5-2
times that of $L$. The ellipticities predicted by this $M\propto L$
model fall below the PSPC data at a significance slightly greater than
the 90\% level. Modeling the gravitating mass with a density run
$\rho\sim r^{-2}$ or with a Hernquist profile we find that the
ellipticity of the gravitating matter is, $\epsilon_{mass}\cong 0.35 -
0.65$ (90\% confidence), which is larger than the intensity weighted
optical ellipticity $\langle\epsilon\rangle = 0.30$.

This evidence for dark matter which is more flattened and more
extended than $L$ is similar to our conclusions from previous X-ray
studies of two other ellipticals, NGC 720 and NGC 1332, but at
somewhat smaller significance level than for NGC 720 (e.g.  Buote \&
Canizares 1997b). These results are consistent with analyses of known
gravitational lenses (e.g.  Keeton, Kochanek, \& Falco 1997), two
polar ring galaxies (Sackett et al. 1994; Sackett \& Pogge 1995), and
flaring disks in spiral galaxies (e.g, Olling 1996). The ellipticities
of the gravitating matter derived from our X-ray analyses and these
other methods are consistent with those of halos produced by CDM
simulations (e.g.  Dubinski 1994).

If an isothermal gas is assumed then models with matter density
$\rho\sim r^{-2}$ are favored over Hernquist models (and similar
models like the universal CDM profile of Navarro et al. 1997).  For
$r\sim 100\arcsec-300\arcsec$ the $\rho\sim r^{-2}$ model marginally
fits the data better than the Hernquist model.  However, most of the
difference in these models occurs in the central radial bins where the
effects of multi-phase cooling flows, magnetic fields, and discrete
sources could affect the surface brightness profiles, though we have
argued the effects are unlikely to be important (see \S \ref{models}).
(The derived shape of the gravitating mass is mostly robust to these
issues -- Buote \& Canizares 1997b.) This support for nearly $r^{-2}$
profiles agrees with previous studies of gravitational lenses (e.g. 
Maoz \& Rix 1993; Kochanek 1995), although a recent paper finds that
density profiles with changing slopes (e.g.  Hernquist and NFW) are
preferred \cite{lilya}.

An emission component that is proportional to $L$ cannot contribute
significantly to the {\sl ROSAT} X-ray emission of NGC 3923, and thus
discrete sources should not affect our constraints on the gravitating
matter (Buote \& Canizares 1997a). However, the {\sl ASCA} spectral
data when fitted with two thermal components yield a cold component,
$T_C=0.55$ keV, and a hot component, $T_H\sim 4$ keV, where the
relative flux of cold-to-hot is $\sim 1.9$ in the {\sl ROSAT} band
\cite{bf}. The conventional interpretation of the hot component (e.g.
Matsumoto et al. 1997; Loewenstein \& Mushotzky 1997) is that it
arises from discrete sources. But our analysis (\S \ref{radpro}) shows
that $\sim 35\%$ of the 0.5-2 keV emission cannot be distributed like
the optical light which would be expected of discrete sources.  Hence,
either the emission from discrete sources is not distributed like $L$,
or the hot component obtained from the spectral fits cannot be
entirely due to discrete sources as suggested by Buote \& Fabian
\shortcite{bf}.

The constraints we have obtained for NGC 720, NGC 1332, and now NGC
3923 from analyses of their X-ray isophote shapes and radial surface
brightness profiles provide an initial demonstration of the power of
X-ray analysis for probing the shape and radial distribution of
gravitating matter in early-type galaxies.  The next generation of
X-ray satellites, particularly {\sl AXAF} and {\sl XMM}, have the
capability to accurately map X-ray isophote shapes and orientations
from the cores ($r\sim 1\arcsec$) out to 10s of kpc for many
galaxies\footnote{The vastly improved spatial resolution of {\sl AXAF}
over the {\sl ROSAT} PSPC will allow easy exclusion of the bright
point source (1) (see Table \ref{tab.src}) which hindered the present
analysis of NGC 3923.}. The spatially resolved spectra provided by
these future missions will allow more precise constraints on
temperature gradients and the contribution from discrete
sources. Thus, unlike most other methods, obtaining interesting X-ray
constraints on the shape and radial density profile of the gravitating
matter will be possible for a large sample of early-type galaxies
since the X-ray analysis is applicable to any isolated early-type
galaxy whose soft X-ray emission ($\sim 0.5-2$ keV) is dominated by
hot gas.

\end{document}